# Thermal Transport in Polymers: A Review

Xingfei Wei[1,#], Zhi Wang[2,#], Zhiting Tian[2,*], Tengfei Luo[1,3,*]

1. Department of Aerospace and Mechanical Engineering, University of Notre Dame, IN, USA

2. Sibley School of Mechanical and Aerospace Engineering, Cornell University, NY, USA

3. Department of Chemical and Biomolecular Engineering, University of Notre Dame, IN, USA

\# -- these authors contributed equally

\* corresponding authors: zhiting@cornell.edu; tluo@nd.edu

**Abstract**

In this article, we review thermal transport in polymers with different morphologies from aligned fibers to bulk amorphous states. We survey early and recent efforts in engineering polymers with high thermal conductivity by fabricating polymers with large-scale molecular alignments. The experimentally realized extremely high thermal conductivity of polymer nanofibers are highlighted, and understanding of thermal transport physics are discussed. We then transition to the discussion of bulk amorphous polymers with an emphasis on the physics of thermal transport and its relation with the conformation of molecular chains in polymers. We also discuss the current understanding of how the chemistry of polymers would influence thermal transport in amorphous polymers and some limited, but important chemistry-structural-property relationships. Lastly, challenges, perspectives and outlook of this field are presented. We hope this review will inspire more fundamental and applied research in the polymer thermal transport field to advance scientific understanding and engineering applications.

1. INTRODUCTION

Achieving high thermal conductivity in polymers is desirable for contemporary applications such as electronics packaging,[1] thermal interface materials [2] and polymeric heat exchangers.[3] These applications have triggered a renewed interest in exploiting thermally conductive polymers using a variety of fabrication methods, such as mechanical stretching,[4, 5] electrospinning,[6] nanoscale templating,[7] compositing,[8, 9] and polymer blending.[10] By drawing amorphous polyethylene (PE) into highly aligned fibers, chain entanglements and voids, which act as stress concentration points and phonon scattering centers in the amorphous form, are significantly reduced, leading to a thermal conductivity increase by two to three orders of magnitude.[4, 11-16]

Besides changing the global morphology of polymers, composites are more commonly pursued in real applications since they are easier to realize and cost-effective. By compositing amorphous polymers with thermally conductive particles, thermal conductivity can be improved to as much as ~10 W/mK,[17-21] and such improvement can be potentially enhanced by strong particle-matrix interfacial adhesion (e.g., covalent bond,[22-24] π- π stacking,[25] and hydrogen bond[26-30]) and better vibrational spectra coupling.[31] Despite consistent improvements in composite thermal conductivity, the bottleneck is still the low thermal conductivity of the polymer matrices. Theoretical calculation has pointed out that a composite thermal conductivity greater than 20 W/mK is more likely to be achieved if the polymer matrix has an intrinsic thermal conductivity larger than 1 W/mK.[32] However, the details of thermal transport in pure amorphous polymers are still not thoroughly understood, leaving limited guidance on designing polymers with intrinsic high thermal conductivity.

Thermal transport in polymers is an active field, and there have been a few reviews on thermal transport in polymers and their composites in recent years.[32-37] Henry's review[32] is 7 years old, and there have been a large number of new studies in recent years. Chen et al's review[33] focused mostly on polymer composites, and the rest of the reviews surveyed both polymers and composites. A few of these reviews are from a more engineering viewpoint (e.g., Refs.[33, 37]). The review from Xu et al.[35] emphasized physical understanding but it is comparatively short. We believe an updated and comprehensive review with the focus on thermal transport physics and chemistry-structure-property relation in pure polymers is needed.

In this article, we strive to deliver a comprehensive review of the state of thermal transport in pure polymers with the emphasis of the above two aspects (i.e., thermal transport physics and chemistry-structure-property relationships). We first review recent advancements in the understanding and engineering of thermal conductivity through changing the global morphology of polymers. Specifically, we will discuss the physical origin of the high thermal conductivity in polymer chains, and how this was realized in experiments. This is followed by more discussion of studies that try to understand factors that influence the thermal conductivity of aligned polymers using mainly molecular dynamics (MD) simulations. We then shift to the discussion of bulk polymers and the recent experimental and modeling studies aiming to reveal the fundamental thermal transport physics. Notable works that produce thermal conductivity higher than 1 W/mK are discussed and the mechanisms also analyzed. We also discuss the concept of using external stimulation to tune the thermal conductivity of polymers. Lastly, we discuss challenges, perspectives and outlook in this field.

## 2. THERMAL TRANSPORT OF ALIGNED POLYMER FIBERS

Polymers, those once believed to be thermal insulators, are now breaking the boundary between insulators and thermal conductors, thanks to a few seminal works showing that the thermal conductivity of polymer fibers can be even higher than common metals.[4, 11, 38, 39] Fibrous or elongated polymers have much higher thermal conductivity when compared to their bulk counterparts. Bulk polymers unavoidably have both compositional and structural defects and disorders, which introduce significant phonon scattering along the heat conduction pathway or even change the nature of heat carriers and their transport mechanism (see Section 3). Thus, polymer fibers/nanofibers, which can be engineered at the molecular level, have come into the light and become a highly promising class of material for thermal conduction applications.

Heat conduction in low-dimensional materials has long been a research topic of interest since it offers an excellent platform for studying fundamental thermal transport theories, where early physics models have paved the way for engineering research.[40-44] Fourier's law has been governing heat conduction in macroscopic materials, but when thermal conductivity becomes size-dependent, Fourier's law fails, and it is known as anomalous heat conduction.[45] It is found that for 1D lattices, when the momentum is conserved, the thermal conductivity is divergent with the length,[46] $L$, in a power law of $\kappa \propto L^{\beta}$.[47] The divergent exponent has been reported to fall between $\frac{1}{3}$ and $\frac{2}{5}$.[45, 48] This anomalous heat conduction was later attributed to anomalous phonon diffusion.[49] While comparing $L$ with the phonon MFPs, we should bear in mind that MFPs have a wide distribution in a given material.[50]

We devote this section to discuss the understanding and advancement of thermal transport in polymers with aligned crystalline morphology, mostly fibers. We start with the discussion of simulation, theoretical and experimental tools used for thermal transport measurement and understanding of polymer fibers, which are also popular tools for bulk amorphous polymers. We then discuss different synthetic methods developed for fabricating polymer fibers with high thermal conductivity. Finally, we focus on how the microscopic and molecular-level structure of the polymer fibers affects phonon transport and thermal conductivity. In polymer fiber studies, PE is the most researched material, which stands out to be one of the most promising polymers to attain high thermal conductivity, and it has been a great platform to understand interesting physics related to phonon transport and structure-property relations. There has also been much work done on polymers like polythiophene (PTs), polyimide (PI), polystyrene (PS), polyurethane (PU), and polymer composite fibers, which will also be mentioned in this section.

## 2.1. SIMULATION METHODS

Computation has led the way of understanding and exploring thermally conductive polymer fibers. There are often two steps in using a model for thermal conductivity calculation.[51-53] The first step is to construct polymer fibers with designated length, structure, and conformation with software like BIOVIA Material Studio. The construction of the materials is vital since a small change in structure can lead to significant changes in thermal conductivity, especially for aligned polymer fibers.[54-56] Usually, careful relaxation and minimization of both the atomic configuration and simulation cell size are needed to ensure reaching reasonable starting structures.[56] Then, simulation methods like MD or density functional theory (DFT) are applied to model the polymer fibers and calculate thermal conductivity. DFT can be more accurate in modeling interatomic interactions since it does not need empirical potentials, but it is computationally expensive and is best suited to model perfect crystals to predict the upper limit of thermal conductivity. MD, on the other hand, can simulate much larger systems to account for factors like defects and amorphous conformation, but its accuracy depends on the fidelity of empirical potentials. MD has been a valuable tool to explore the structure-property relations of polymers. Each method is further discussed in the following sections.

### 2.1.1. DFT Calculations

The first-principles anharmonic lattice dynamics calculation is based on the computation of the interatomic force constants extracted from DFT. First-principles-based lattice dynamics predicted that the thermal conductivity of an individual PE chain could be as high as 1400 W/mK,[57] while that of a 100 nm crystalline PE fiber can reach 310 W/mK.[58] With the force constants from DFT calculations and relaxation times calculated from Fermi's Golden rule, one can calculate thermal conductivity by solving the phonon Boltzmann Transport Equation (BTE):[59, 60]

$$k = \frac{1}{V} \Sigma_P \Sigma_k C_{p,k} v_{p,k}^2 \tau_{p,k} \tag{1}$$

where $C$ is the specific heat per mode, $v$ is the phonon group velocity, $p$ is polarization, $k$ is wave vector, $V$ is the volume, and $\tau$ is the relaxation time.

Temperature-dependent effective potential methods (TDEP) have been applied in the DFT calculation for thermal conductivity to help reduce computational expense when comparing to MD.[60, 61] In the TDEP, atoms in a supercell of the crystal have thermal amplitudes,[60] which correspond to a canonical ensemble of the target temperature. TDEP includes zero-point motion and finite-temperature anharmonicity, which is accomplished by sampling the Born-Oppenheimer energy surface at a designated temperature. This sampling includes additional displacements from quantum nuclear motion. Effective harmonic and cubic interatomic force constants can be used with the anharmonic perturbation theory to calculate thermal conductivity.

### 2.1.2. MD SIMULATIONS

MD simulation is one of the most widely used and powerful tools for thermal conductivity calculation of polymer fibers.[39, 51-53, 62, 63] Using this method, Henry et al. predicted that a single extended PE chain would have infinite thermal conductivity due to lack of ergodicity,[39] which largely re-fueled the study of thermally conductive polymer fibers experimentally.[4] MD simulations can model much larger structures (up to millions of atoms) than DFT, which helps research reveal the structure-property relation that is more relevant to real conditions. It also allows the study of thermal transport physics related to phonon scattering of defects (e.g., segmental rotation, amorphous regions, chain ends, or voids) – important factors preventing polymer fibers from reaching the theoretical limit from DFT calculation of perfect crystals.[54-56, 64] There are two different kinds of MD methods used for thermal conductivity calculations, equilibrium MD (EMD) and non-equilibrium MD (NEMD), where EMD is conducted in an equilibrium state without temperature gradients, while in NEMD, a steady-state temperature gradient is established. A typical MD simulation consists of three steps: (1) starting structure construction; (2) relaxation and optimization; and (3) production runs. A reasonable starting structure followed by careful relaxation and optimization is critical to achieving the correct molecular configuration to accurately predict the thermal conductivity, which can be a strong function of the subtle conformation of chains.[12, 54-56, 65, 66]

EMD is usually combined with the Green-Kubo formula for thermal conductivity calculation:[39, 55, 67]

$$k_x(T) = \frac{1}{Vk_BT^2} \int_0^\infty \langle \vec{J}_x(0) \cdot \vec{J}_x(\tau) \rangle d\tau \qquad (2)$$

where $k_x(T)$ is thermal conductivity along the polymer chain direction, $V$ is volume, $k_B$ is Boltzmann constant, $T$ is temperature, $\vec{J}_x$ is the heat flux in the x-direction, $\tau$ is the correlation delay time. EMD for polymers usually needs a large ensemble average to obtain converged thermal conductivity. The often-used periodic boundary condition in EMD can potentially lead to an artificial correlation between certain phonon modes, which might be especially true for crystalline polymer fibers with high thermal conductivity. However, this effect has not been fully understood nor quantified.

In NEMD methods, either a temperature gradient is established via thermostats (i.e., normal NEMD), or a scheme called reverse NEMD[68] is used to swap atom kinetic energy between the two thermal reservoirs to establish a heat flux. When the system reaches a steady-state, the thermal conductivity is calculated by Fourier's law, $k = -\frac{J}{\frac{dT}{dx}}$, where $J$ is heat flux, and $\frac{dT}{dx}$ is temperature gradient in the heat flux direction. For polymer thermal conductivity calculation, NEMD usually predicts reproducible data without the need for large ensemble averaging, and there is a better-understood size effect that can be handled via extrapolation.[69, 70]

The choice for heat bath is crucial for calculating accurate thermal transport properties in MD simulations, especially for systems where long wavelength phonons dominates.[71, 72] Two frequently used heat baths are Nosé-Hoover heat bath (deterministic and time-reversible) and Langevin heat bath (stochastic and time-irreversible).[71] The generation and accumulation of localized edge modes in Nosé-Hoover bath require multiple layers of Nosé-Hoover bath to diminish the boundary layer's temperature differences.[73, 74] In contrast, Langevin baths have stochastic excitation of all modes, which prevents the accumulation of localized edge modes and generates marginal temperature differences at the boundary. Usually, we find that amorphous polymers are more immune to the effect of thermostats since the structural disorder can quickly scatter phonons generated by the thermostats to recover the intrinsic phonon distribution. However, a recent study showed that heat carrier MFP in amorphous polymer can be tens or even more than 100 nm, and there can be size effect in NEMD simulations.[75] As a result, the effect of thermostats in NEMD calculation of amorphous polymer thermal conductivity deserves a detailed study in the future.

The accurate prediction of thermal conductivity depends on the fidelity of the interatomic potentials. Both all-atom and united atom models have been used for polymer thermal conductivity calculations.[39, 63, 67, 76] The major difference between these two types of force fields is whether hydrogen atoms are explicitly represented. Henry et al. argued that the explicit simulation of hydrogen atoms is necessary to capture all degrees of freedom, which is important to thermal conductivity. However, the comparison was between two different potentials (Adaptive intermolecular reactive bond order (AIREBO)[77] vs. Kirkwood[78]). It is not yet conclusive on how important an effect the united atom model has on thermal conductivity. However, the hydrogen degrees of freedom are largely un-excited at room temperature due to quantum effects – a justification used in many studies using the united atom models. A few most commonly used all-atom potentials for thermal conductivity studies include AIREBO potential,[77] the Class-II potential PCFF (polymer constant force field),[79] and COMPASS force field (Condensed phase optimized molecular potentials for atomistic simulation studies).[80-82]

## 2.2. Theoretical Model

Since both DFT and MD can be computationally expensive, a model that can predict thermal conductivity would be valuable. A simple model has been proposed for this purpose for crystalline polymers, where only a unit cell structure is needed for the prediction:[83]

$$k_{TM} = A \frac{\bar{E}^{\frac{2}{3}}}{\overline{M_b}^{\frac{1}{2}} V^{\frac{2}{3}} T} P^a \delta^b \left(\frac{\overline{M_s}}{\overline{M_b}}\right)^c \quad (3)$$

where $A$, $a$, $b$ and $c$ are coefficients that can be fitted to crystalline polymer data, $P$ stands for chain rotation ratio, $\delta$ is in-plane bond ratio, $\overline{M_s}$ and $\overline{M_b}$ are atomic mass, $\bar{E}$ denotes bond energy, $V$ is volume of the unit cell and $T$ is temperature. The establishment of this model may facilitate researchers to perform preliminary thermal conductivity screening for crystalline polymers and spares them the efforts to undergo DFT or MD simulations. Besides the calculation efficiency, this model also displayed high accuracy in comparison with data from more expensive simulation methods (**Fig. 1**).

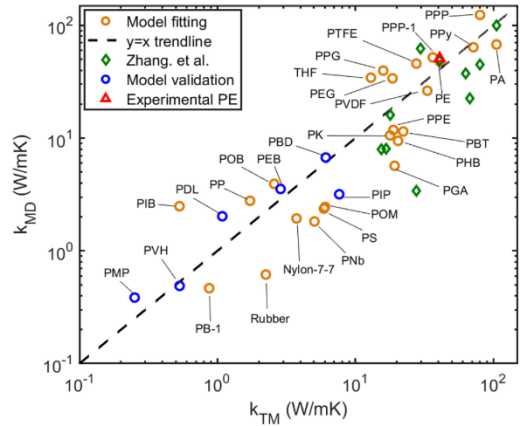

**Figure 1.** A simple model for predicting thermal conductivity of polymer crystals and its comparison to other MD simulation results. [83] Reproduced with permission.[83] Copyright 2019, American Chemical Society.

## 2.3. THERMAL CONDUCTIVITY MEASUREMENT TECHNIQUES FOR POLYMER FIBERS

Unlike bulk polymer materials, measuring the thermal conductivity of polymer fibers requires unique setups and equipment with high precision and fine spatial resolution. Several techniques, such as the thermal bridge method,[84] atomic force microscopy (AFM) cantilever assisted measurement, and time-domain thermoreflectance (TDTR), have been employed for measuring polymer fiber thermal conductivity. **Table 1** shows the comparison of these different measurement techniques, which are further discussed below.

**Table 1.** Comparison between different measuring techniques for the thermal conductivity of polymer fibers.

|  | Thermal bridge method[7, 11, 12, 85-87] | *AFM cantilever method*[88] | *TDTR and other transient methods*[16] |
|---|---|---|---|
| *Advantages* | Simple setup, high precision | Highest precision | High precision |
| *Disadvantages* | Manual fiber placement is difficult | Cannot measure stiff samples, complex setup | Cannot measure fibers with diameters smaller |

**Thermal bridge method:** Thermal bridge method (**Fig. 2a**) is straightforward with relatively high precision based on Fourier's law for measuring thermal conductivity values.[7, 11, 12, 85-87] It was used more than two decades ago when people directly attached the polymer fiber to a heat sink and a heater for the steady-state heat flow.[87] However, precision and sample size requirements were largely limited at that time, where only micron-sized fiber or fiber bundles could be handled. The thermal bridge method has been commercialized and standardized, which allows for the measurement of much thinner fibers (down to dozens of nanometers) with improved precision. The modern thermal bridge method consists of two separate islands of heaters/sensors made of platinum/$SiN_x$ maintained at different temperatures. The polymer fiber serves as a bridge connecting these two islands and conducting heat from the high to the low-temperature islands. A combined DC and AC current is introduced to one island for Joule heating and resistance measuring, making this island the heater component. Simultaneously, on the opposite island, an AC with the same value is applied for sensing the resistance across the bridge. The thermal conductance of the fiber can then be measured.

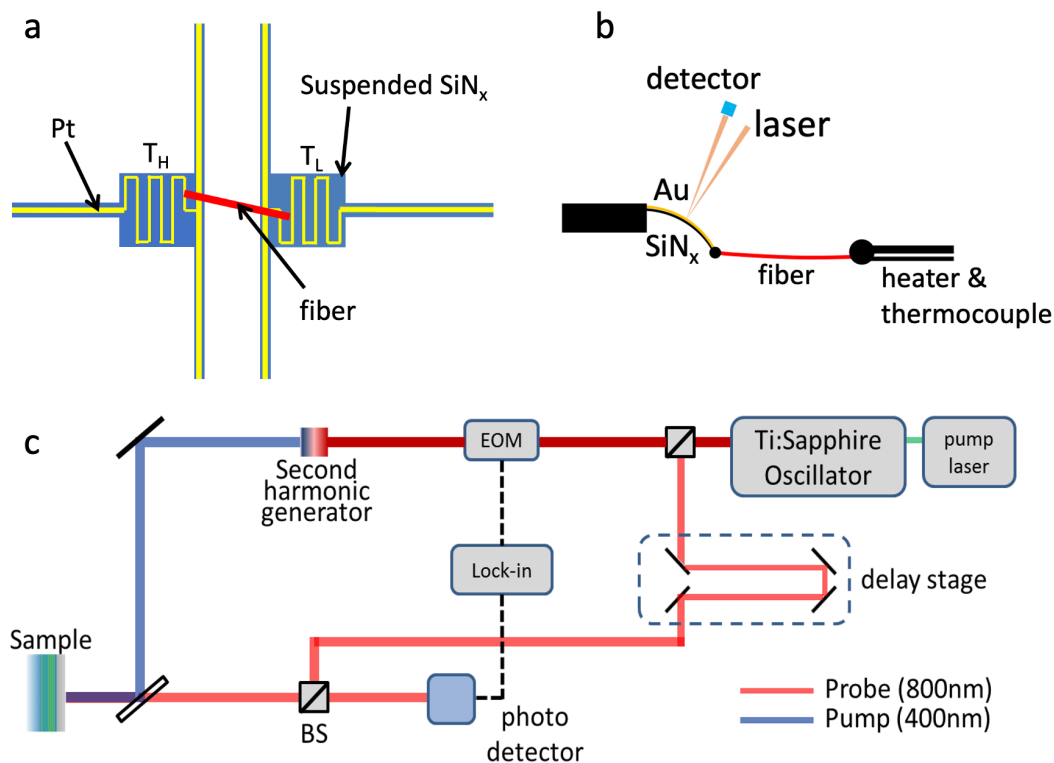

**Figure 2.** Schematic of (a) thermal bridge method, (b) an AFM cantilever measurement and (c) a TDTR system.

**AFM cantilever method:** AFM cantilever-assisted measurements, usually performed in high vacuum, is suitable for measuring thin fibers with nanosized diameters, and this method has a very high precision since it excludes much of the influences such as heat loss, thermal expansion, and thermal contact resistance. In a typical AFM cantilever method setup (**Fig. 2b**), a tipless AFM cantilever bi-material ($SiN_x$ coated with a gold film) is used for drawing and connecting one end of the polymer fiber.[88] The other end of the fiber adheres to a thermocouple on a needle tip made of conductive silver epoxy. Electrical current was applied to the needle for Joule heating and temperature adjustment. A laser beam with a wavelength of 650 nm was focused onto the tip of the AFM cantilever, which is subsequently reflected onto a receiving photodiode for deflection measurement. As the temperature of the AFM cantilever changes, it bends due to the different thermal expansion coefficients of two materials ($SiN_x$ and Au), and the amount of bending, detected by the laser deflection, indicates the cantilever temperature.

**TDTR method:** TDTR (time-domain thermoreflectance) can measure both the axial and radial thermal conductivity of the polymer fiber.[16, 89] A laser is used as a power source for producing laser pulses with a constant repetitive frequency. A beam splitter will divide the laser into a pump and a probe beam whose optical paths are adjusted by a series of delay stages. Finally, both the probe and pump beams will be focused by an objective lens onto the surface of the sample (**Fig. 2c**). A multi-layer heat transfer model is fitted to obtain the thermal conductivity from TDTR signals. When measuring fiber samples, the fibers are first embedded into epoxy and then cut open to expose the cross-section of the fiber. A metal transducer layer is then deposited on the exposed cross-section, and TDTR is performed so that the thermal conductivity in the fiber length direction is measured. However, the application of this method is mostly limited by the laser spot size, which should be smaller than the cross-sectional area of the fiber. Thinner fibers require an objective lens with a higher magnification ratio, and most of the measuring samples are thicker than 10 μm. To measure the heat conduction along the radial direction, the polymer fibers need to be indented to create a flat surface before the deposition of the transducer layer.

**Calorimetric scanning thermal microscopy (C-SThM):** All the techniques mentioned above have limitations when it comes to fibers with extremely small diameters, such as below 10 nm. C-SThM may be a potential candidate to measure very thin fibers because it is possible to measure the thermal conductance of single molecules.[90] A SThM tip serves as a heater and the substrate for the molecule as the heat sink, and both of them are made of gold. Between them is the molecule (e.g., a single alkanedithiol molecule) subject to measurement. For sample preparation, the gold substrate was immersed in a diluted solution of target molecules for self-assembly, then the molecule would form a bridge between the SThM tip and the gold by moving the SThM tip away from the gold substrate. C-SThM represents some of the most delicate systems to measure extremely small thermal signals, which makes measuring single molecular thermal conductance possible.

## 2.4. POLYMER FIBER SYNTHESIS METHOD

In pursuit of higher thermal conductivity, researchers had come up with many different synthesis techniques for manufacturing polymer fiber/nanofiber. Spinning is the most mentioned method for making polymer fibers in mass production, and ultra-drawing has been effective for further increasing the thermal conductivity. For spinning, there are mainly wet spinning, dry spinning, electrospinning, melt spinning, and gel spinning, which are more suitable for large-scale manufacturing at low cost.[91] **Table 2** compares these spinning methods. Among them, electrospinning is the most researched method because of its relatively small batch-to-batch difference and easy-to-control nature, which has been applied to polymers such as PE,[92] PI,[93] Nylon,[6] PAN, and PMMA.[94] There are also other synthesis methods for polymer fibers. For example, the nanotemplate can produce an array of polymer fibers with high thermal conductivity.[95] Utilizing a nano-porous nanotemplate, it is possible to obtain chain-oriented polymer nanofibers. **Figure 3a** shows a comparison of thermal conductivity among different synthesis methods. Fiber drawing is both experimentally[38, 88] and theoretically[14, 54] proven to be effective for improving thermal conductivity, and experimental results have shown that it is possible to achieve thermal conductivity over 100 W/mK,[88] which makes it comparable to that of many metals. Heat stretching after the tip-drawn step can further increase the thermal conductivity.[11] In the following sub-sections, we briefly discuss the synthesis methods used to achieve high thermal conductivity polymer fibers.

Table 2. Comparison between different spinning methods used in thermal conductivity studies

|  | *Status of the pre-fiber polymer* | *Spinning method* | *Working temperature* | *References* |
|---|---|---|---|---|
| *Electrospinning* | In solvent | Electric field | Varies | 86, 92-94, 96, 97 |
| *Wet spinning* | In solvent | Simple drawing | Room Temperature | 98 |
| *Dry spinning* | In solvent | Mechanical extrusion | Hot air drying | 99 |
| *Melt spinning* | *Melt polymer* | *Mechanical extrusion* | Above melting temperature | 100 |
| *Gel spinning* | *Gel state* | *Mechanical extrusion* | In melting range | 101 |

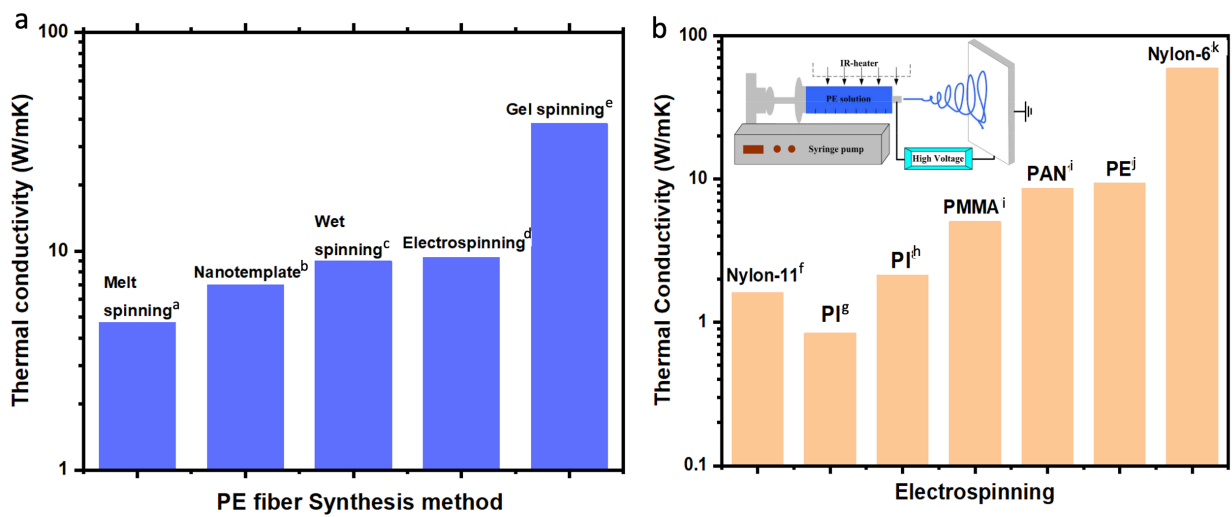

**Figure 3.** (a) Comparison of thermal conductivity of PE fibers synthesized from different methods. Melt spinning: extrusion ratio of 54; Nanotemplate: 200 nm in diameter for HDPE or 100 nm in diameter for P3HT; Wet spinning: 40 nm in diameter; Electrospinning: 53 nm in diameter; Gel spinning: a large fraction of long (>50nm) extended chain crystal in the fiber (b) Comparison of thermal conductivity of different polymer fibers from electrospinning. Inset: schematic of a typical electrospinning setup.[92] Reproduced with permission.[92] Copyright 2009, RSC Pub. Refs: a-[100]; b-[95]; c-[98]; d-[92]; e-[101]; f-[6]; g-[86]; h-[93]; i-[94]; j-[92]; k-[97].

**Electrospinning:** Electrospinning is one of the most popular methods in the large-scale production of polymer fibers. A typical electrospinning setup is often composed of a heated syringe pump, a high voltage region (> 10 kV) for spinning, and a grounded fiber collector (inset in **Fig. 3b**).[92] Under this high voltage electric field, the polymer solution will form a Taylor cone and generate an elongated jet flow, becoming progressively thinner before reaching the grounded collector because of solvent evaporation and external force field. Besides its low cost, electrospinning has the advantage of being suitable for a wide range of polymers from high melting point polymers like PI[93] to lower ones like PE,[92] or even composite polymer fibers like PAN/graphene or PMMA/graphene composite fibers.[94]

Processing conditions in electrospinning are essential for obtaining high thermal conductivity for polymer fibers.[92] A higher voltage will lead to a stronger working electric field for the spinning process, increasing the elongation force and ratio and increasing the chain alignment in the fibers.[6, 86] This is in line with the finding that the stretching process during spinning will reduce the fiber diameter, and fibers with smaller diameters will improve the orientation of crystallites in the polymer fibers, which in turn increases the thermal conductivity.[98] However, too high a voltage will reduce the flight time for the spun polymer fibers to grow crystalline areas. Due to the existence of whipping instability in the electrospinning process, the thermal conductivity of the as-spun fibers can vary significantly even if other conditions are the same.

**Solid-state extrusion (melt spinning):** A solid-state extrusion setup comprises a piston that exerts a high pressure onto the polymer, a cylindrical cavity that serves as a container for the raw materials, and a die that gives shape for to-be-extruded polymer fibers.[100] This technique is a simplified version of electrospinning. Solid-state extrusion products often come with a high extrusion rate, and it improves the thermal conductivity of polymer fiber by increasing crystallinity and better chain and lamellae orientation, both of which benefit phonon transport along the chain direction.[100]

**Gel-spinning:** Gel-spinning[101] (also called dry-wet spinning) is a method very similar to solid-state extrusion, but the polymers are under a gel-state (partly liquid and partly solid). For gel-spinning, polymers go through several steps before becoming fibers. First, the polymers are dissolved in an organic solvent at elevated temperatures. This solution is then extruded and goes through a two-stage cooling and drying, where the air will remove excess solvent, and water cooling will quench the polymer into a gel state. Finally, spinneret will spin the gel state polymers into fibers. In gel-spinning, molecules will intertwine with each other, and polymer chains in the gel state will bind together and produce an enhanced inter-chain interaction along the chain direction, forming a network inside the polymer fiber and making it easier to obtain fibers with a higher draw ratio. This is the reason why gel-spinning tends to produce the highest thermal conductivity among all these other methods.

**Wet spinning (Tip drawing)**: Wet spinning is the oldest method to produce polymer fibers from bulk material. It does not need any special equipment, and only a polymer solution is needed for fiber fabrication. A typical method for wet spinning is tip spinning, where a needle tip dips into a polymer solution at an elevated temperature, and as it pulls out, a fiber can be drawn from the solution.[98] Since the drawing process is at room temperature, more defects tend to emerge in the drawn fiber. If the drawing rate is not precisely controlled, the room temperature cools down the fiber quickly, and the molecular chains will not have enough time to rearrange themselves. This method is not easily scalable for manufacturing polymer fibers in mass production.

**Dry spinning:** Dry spinning is a method developed for heat-sensitive polymer fibers.[99] Typical dry spinning equipment is composed of three parts, including a polymer container with a spinneret, which extrudes polymer fibers, an evaporation cabinet for solvent evaporation, and a stretching device. In dry spinning, bulk polymers are first dissolved in volatile organic solvents to form a low viscosity fluid, which is then extruded into an evaporation cabinet where hot inert gas like nitrogen is used to evaporate the solvents in the polymer fiber. Finally, the dried fiber will be collected and further drawn to achieve a higher fiber orientation. It is a handy method for heat-sensitive polymers, polymers susceptible to thermal decomposition, such as polymers like PVC (polyvinyl chloride), cellulose acetate, and polybenzimidazole (PBI).[99] However, the need for volatile solvents to some extent limits the selection of processable polymers.

**Nanotemplate:** Apart from spinning, nanotemplate, which is convenient and features small batch-to-batch variation, is another method for large-scale manufacturing of high thermal conductivity polymer fibers.[7, 95, 102] Nanotemplate can use commercial AAO (nanoporous aluminum oxide) templates, and it has been shown that the as-synthesized polymer fiber can be easily separated from the template by etching it away, which does not damage the polymer fibers or change their structures.

A typical nanotemplate method comprises three steps.[95] First, polymers are squeezed into an empty AAO nanotemplate under heat and pressure; then transfer the template into a heated vacuum environment to allow for polymer infiltration into the nanotemplate; the final step is dissolving the template with a strong alkaline solvent. Thermal conductivity results from poly(3-hexylthiophene-2,5-diyl) and non-conjugated

PE nanofibers synthesized from melt-processed nanotemplate method showed that a smaller fiber diameter and higher molecular weight would yield higher thermal conductivity (7 W/mK). [95] It is also feasible to begin this process with monomers, and methods like electropolymerization were employed to polymerize the monomers to synthesize chain-oriented PI fibers.[7] The resultant PI fiber had a thermal conductivity of 4.4 W/mK. A combination of high thermal conductivity and high-temperature thermal stability (200 °C) makes it a possible candidate for TIM (thermal interfacial material) in electronic devices,[7] but the stiffness of the film is a concern for forming conformal contacts with rough surfaces. Different from spinning, nanotemplate may be used to synthesize large arrays of polymer fibers, but using AAO template as a consumable for such fabrication makes the cost a major concern. In addition, the shape and shape distribution of the nanofibers largely depend on the quality of nanotemplate. Even a small difference in these templates will lead to a massive difference in thermal conductivity.[7, 103]

**Other methods:** Other methods used to produce thermally conductive polymers include oCVD (oxidative Chemical Vapor Deposition)[104] and solvent separation.[105] oCVD is a one-pot reaction method, where all the precursors are gathered as vapor phases in a vacuum chamber. Heating and pressure conditions are tuned to trigger the step polymerization reaction. The advantages of this method are that this is an all-dry species vapor phase reaction, where no solvent is presented, which allows for easier fabrication of films without the need for post-processing. Moreover, oCVD makes it possible to take control of both intramolecular and intermolecular structures when polymerization happens. Solvent separation for polymer fiber fabrication is comparatively more complex, and it only works for certain types of polymers (e.g., PA[105]) with limited yields. This method can be useful for small batch lab synthesis when spinning or other types of synthesis are not available.

## 2.5. DRAWING-INDUCED HIGH THERMAL CONDUCTIVITY NEAT POLYMER FIBERS

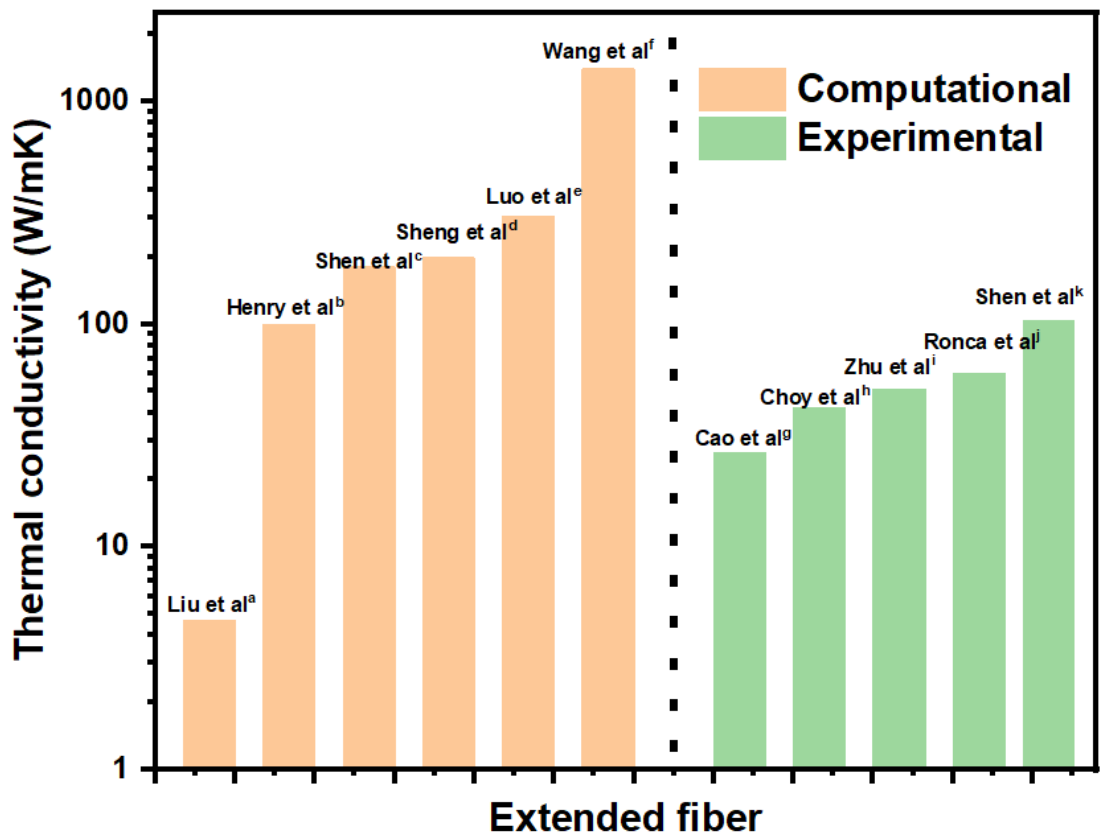

**Figure 4.** Comparison for thermal conductivity of extended polymer fibers between computational value and experimental. Features for a to k are single strand PE, PE longer than 40 nm, PE nanofiber, PTs chain, infinite long PE, single individual PE chain, HDPE nanowire, ultra-drawn PE, heat stretched PE, UHMWPE with a drawn ratio of 240 and PE nanofiber, respectively. Refs: a-[100]; b-[95]; c-[98]; d-[92]; e-[101]; f-[101]; g-[106]; h-[93]; i-[107]; j-[108]; k-[97].

Polymer fiber drawing (also called elongation) can significantly improve the thermal conductivity.[38, 88, 109, 110] **Figure 4** shows the thermal conductivity of extended fibers from computation and experiments. It was proven that under a high drawing ratio and a slow drawing rate, one could obtain extremely high thermal conductivity.[88] In stark contrast to their bulk counterparts, polymer fibers with drawing ratios greater than 400 can have more than two orders of magnitude higher thermal conductivity. Large elongation can effectively alter the orientation, distribution, and morphology of molecular, chain segments, crystalline and amorphous areas, and microscopic structures.[11, 14, 98, 107] A combination of optimal drawing rate and drawing ratio largely determines the amplitude of thermal conductivity for a polymer fiber.[107]

During drawing, when the strain on polymer fibers increases, contribution to thermal conductivity from bonded interaction dominates, and nonbonded interaction like vdW becomes a minor factor.[76] Thermal transport is much more efficient along the covalent-bonded backbone than through vdW interaction.[111] Stretching the polymer fiber by a higher ratio makes the chains more orientated along the drawing direction, taking more advantage of the strong covalent polymer backbone for thermal transport.[66] Besides inefficiently transferring heat themselves, inter-chain vdW interactions can scatter phonons transport along the molecular backbone, thus lowering the thermal conductivity.[62]

The significant increase in polymer fiber thermal conductivity after the drawing is intimately related to the change of the morphologies in the fiber structure. Under a strong strain field, twinning will facilitate the lamellar structure changing into a fibrillar structure.[110] Compared to the lamellar structure, these microfibrillar structures are domains with bundles of highly ordered and oriented chains along the heat conduction direction, facilitating phonon transport and lowering their scattering. [11, 100] There are two stages of structural change in mechanical stretching, which are crucial in understanding the effects on thermal conductivity:[101] (1) The amorphous area will form microfibrils. The crystalline lamellae area will shatter into several smaller crystalline blocks sandwiched by those amorphous areas; (2) Some intrafibrillar tie molecules will connect the crystalline blocks, which lie outside the microfibrils region in the first step. Upon larger strains, these intrafibrillar molecules will extend and align themselves. A higher strain ratio and slower strain rate can both improve the chain orientation in polymer nanofibers.[109] When the polymer fiber is stretched under a much slower elongation process, there will be more time for the uncoiling of chain segments and rotation of chains to adjust themselves for better alignment. However, if the polymer fiber is not perfect in radius or chain segment distribution, excessive drawing can lead to the rupture of nanofiber inside the polymer fiber, which can also contribute to phonon scattering. Furthermore, a drawing speed too high can also be detrimental for thermal conductivity improvement, which will lead to internal stress buildup and limit the achievable elongation ratio at a given temperature.[76, 112]

Heat stretching could further improve thermal conductivity of polymer fibers. During heat stretching, heat allows more movement of chain segments, giving them additional time to rearrange themselves when the crystallization process is not enough for chain alignment. This will increase orientation of crystallites, increases chain alignment, lowers entanglement of chain segments (inset in Fig. 5a, 5c), and reduces chain-chain spacing and configuration disorder of amorphous domains in the fibers.[6, 11] Both local heating and regular bulk heating are processes that involve heating part/whole section of the fiber above the glass transition temperature but lower than their melting temperature. The local heating method was found to be advantageous as it can achieve higher strain rate and minimize the relaxation of the molecules.[11]

In terms of how heat stretching influences fiber crystallinity, however, there have been mixed messages. While local heating was found to increase the crystallinity,[11] other reports showed that heat stretching could lower the crystallinity.[84,89] It seems the different processing conditions may play a role in the change of crystallinity during heat stretching. In the case of reduced crystallinity (Fig. 5b), the enhancement in thermal conductivity was attributed to the shrinking of the amorphous region, and the more ordered structure in the remaining amorphous regions.

There has been much work done on polymer fibers' axial direction since it has the best potential to yield high thermal conductivity. However, recent work done by experiments[89] and simulations[113] started to explore the radial thermal conductivity in polymer fibers. Under mechanical strain, the phonon focusing effect is the main reason for a decrease in thermal conductivity in the radial direction, where the phonon velocity increases along the axial direction and decreases in the radial direction. Without strain, MD simulations showed a strong size effect in the radial direction for thermal conductivity, and the thermal conductivity in the radial direction might be much higher than their amorphous counterparts.[113]

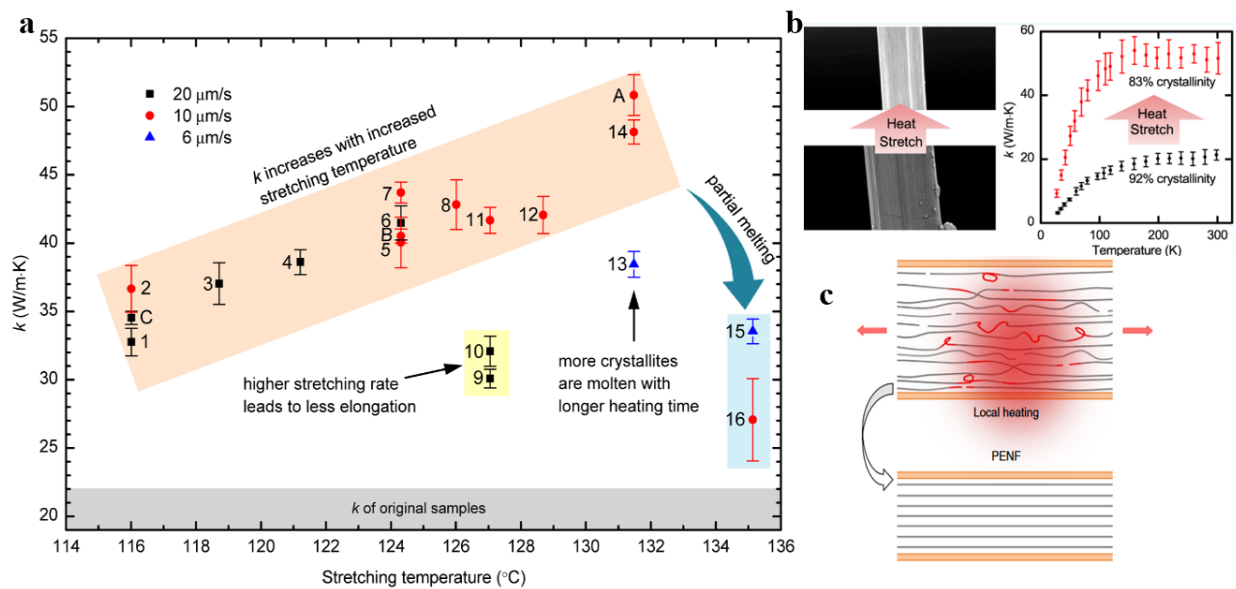

**Figure 5.** (a) Influence of stretching temperature, stretching speed and heating effect on thermal conductivity of elongated polymer fibers;[84] (b) Effect on crystallinity and thermal conductivity after heat stretch;[107] Reproduced with permission. (ref. 84) Copyright 2017, American Chemical Society (c) Uncoiling of polymer segment in local heating process.[11] Reproduced with permission.[11] Copyright 2018, Springer Nature.

## 2.6. POLYMER FIBER COMPOSITES

Like bulk polymer materials, incorporating thermally conductive fillers can also potentially enhance the thermal conductivity of polymer fibers. Studies have reported that spinning is viable for fabricating fibrous nanocomposites, which exhibit much improved thermal conductivity.[94, 95, 114-116] There are two ways of incorporating nanoparticles into a polymer fiber: nano-bridging and fibrous nanocomposite. Nanoparticles with higher thermal conductivity may form interconnecting networks to enable heat-conducting pathways inside the polymer fibers. For fibrous nanocomposites, the thermal conductivity enhancement mechanism is very similar to that of bulk composites, where the rule of mixture can explain the increasing thermal conductivity[117] and percolation in polymer accounts for the discontinuous increase

Thermal Transport in Polymers: A Review.    21

in thermal conductivity.[116] **Figure 6** shows the thermal conductivity of a list of polymer fiber composites.

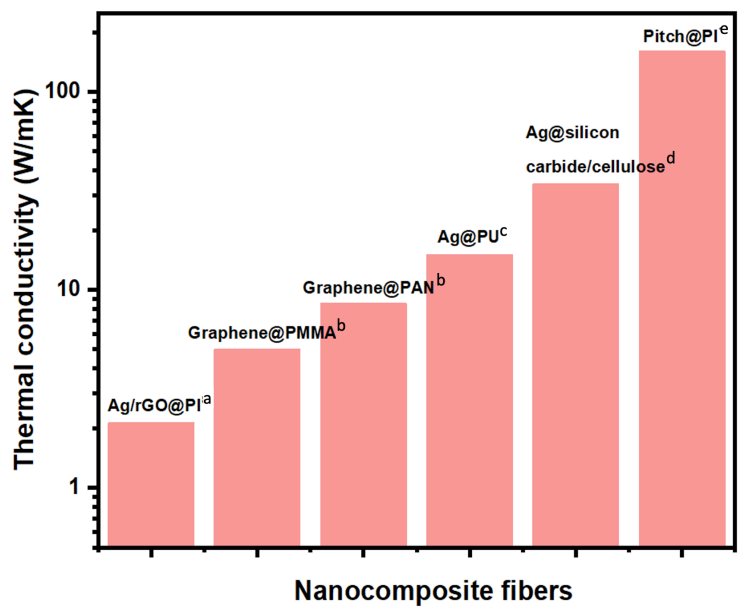

**Figure 6.** Comparison of thermal conductivity between different polymer composite fibers. Refs.: a-[93]; b-[94]; c-[114]; d-[115]; e-[116].

For nano-bridging effects, one of the most used materials is silver.[114, 118, 118, 119] In these composites, small conducting particles bridge the space between long fibers. The high thermal conductivity of silver nanoparticles shows promises in nano-bridging for different kinds of polymer fibers/films. The bridged polymer fibers/films such as Ag/rGO@PI fiber, Ag@silicon carbide/cellulose film, Ag@PU fiber, and Ag/BNNS@Epoxy showed thermal conductivity orders of magnitudes higher than their unbridged counterparts. High filler contents of silver nanoparticles in general lead to decreased inter-filler distance and even a percolation network in the polymer matrix, which can facilitate thermal transport in polymer fiber.

Fabrication of fibrous nanocomposite is a process very similar to bulk nanocomposites.[94, 116] Graphene@PAN fiber,[94] Graphene@PMMA fiber,[94] and Pitch@PI fiber[116] nanocomposites were successfully synthesized. Notably, the percolation network formed by 36 wt% pitch inside a PI polymer fiber introduced 700 times enhancement in thermal conductivity compared to plain PI polymer fibers. It was reported that multiple interpenetrating percolation networks of fillers formed throughout the PI fiber.[116]

It was argued that higher loading of nanofillers could also lead to larger overall interfacial thermal resistance and disrupt the lattice order of the matrix, which could reduce thermal conductivity, suggesting the need to balance these competing effects to optimize thermal conductivity of polymer composite fibers.[105] A recent study on drawn graphene/PE films, however, showed that adding graphene into PE could facilitate crystallization of the PE matrix.[120]

Polymer composites can be drawn to achieve higher matrix crystallinity. However, it was reported that for composite polymer fibers, too large a drawing ratio would lead to a decrease of filler content per volume, which leads to loss of high thermal conductivity fillers and a decrease of overall thermal conductivity.[114] In drawing Ag/PU nanofibers with small strains, the average distance among embedded silver nanoparticles will decrease, which increases the overall thermal conductivity by forming more heat-conducting pathways. For a larger strain, the filler concentration will decrease in the nanofiber, which cuts off the pre-formed pathway and lowers the overall thermal conductivity.

## 2.7. THERMAL TRANSPORT MECHANISM IN POLYMER FIBERS

### 2.7.1. TEMPERATURE DEPENDENCY

The thermal conductivity of highly aligned polymer fibers depends on temperature (**Fig. 7**). For most polymer fibers, which behave much like crystalline materials, high temperatures will lead to a reduction in thermal conductivity due to increased phonon-phonon scattering, and such a drop accelerates when phase change happens.[12, 66] The disordered phase seriously scatter phonons and thus reduce thermal conductivity significantly compared to the crystalline phase. Thus, the application of high thermal conductivity polymer fibers will be limited by their phase change temperature (i.e., glass transition, $T_g$, or melting temperatures), at which structural change in the polymer greatly influences the thermal conductivity. Below $T_g$ the polymer is in the glass state, where the polymer chain segments are confined for their movement. Above $T_g$, the polymer enters the rubbery state, where molecular chains move more freely and segmental rotation becomes more frequent. This segmental rotation presents structural disorder to scatter phonon transport along the chains, leading to a drop in the phonon mean free path (MFP).[13, 65, 66, 121, 122] There have been a number of polymer fibers (e.g., PE,[13] Nylon,[56] polylactic acid (PLLA)[123]) exhibiting sharp decreases in thermal conductivity related to this temperature-induced effect.

When the temperature is low, phonon propagation inside the crystalline region is limited by the scattering at interfaces or boundaries (size effect), like that in inorganic crystals, and thermal conductivity increases with increasing phonon population.[11, 56, 123] An increase in temperature in this region will not change the polymer structure significantly, while it will increase the heat capacity, which can explain the increasing trend at a lower temperature if we consider Eq. 1.

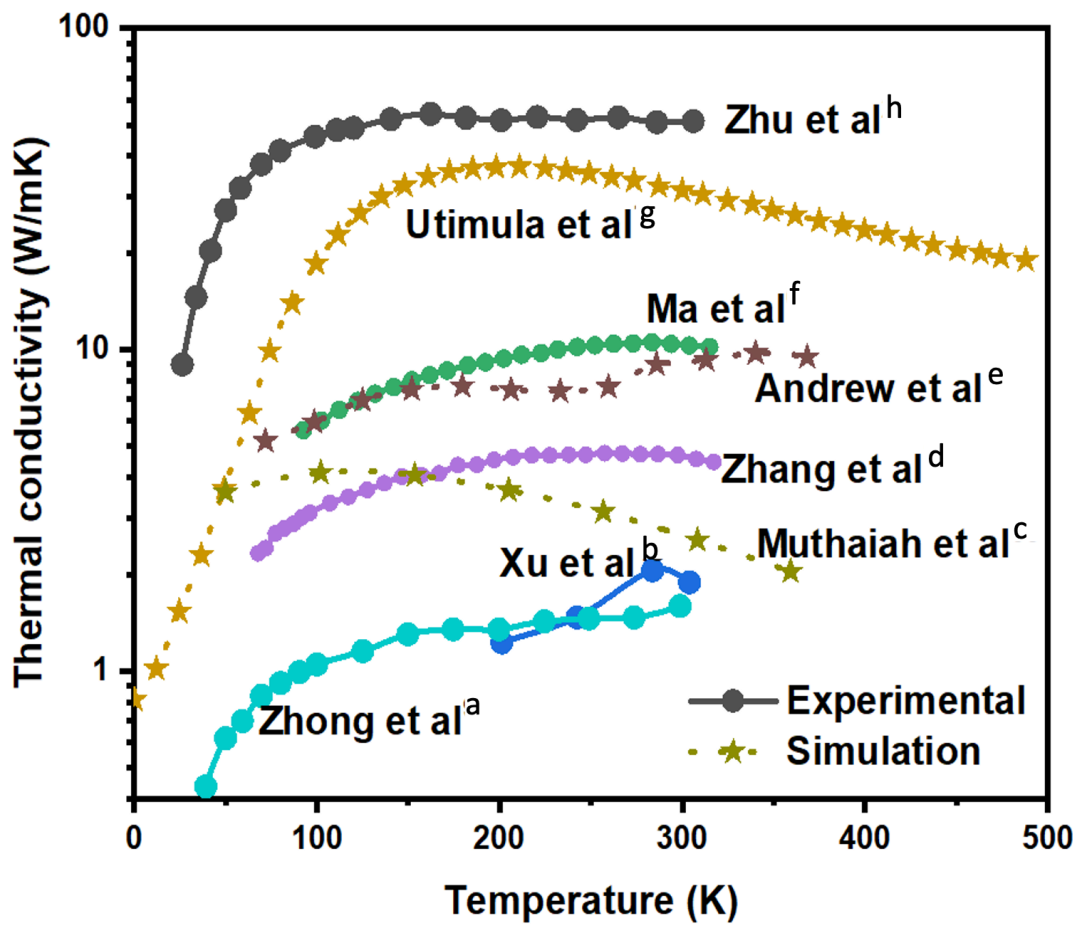

**Figure 7.** Effects of temperature on thermal conductivity of PE fiber. *Refs.: a-[6]; b-[110]; c-[124]; d-[125]; e-[126]; f-[92]; g-[59]; h-[112].*

### 2.7.2. IMPACT OF CRYSTALLINITY, AMORPHOUS PHASE AND THEIR ORIENTATION

Large crystallites and better alignment will increase the intermediate-range order, and phonon MFP, which increases the thermal conductivity.[6] We can often find high crystallinity and large crystallite sizes in most high thermal conductivity polymer fibers. The reason that one can never achieve the theoretical value of high thermal conductivity is that, in reality, it is impossible to synthesize purely crystalline polymers. However, it is shown that factors like orientation, crystalline-amorphous interfaces, and phase distribution need to be considered, and thus a high crystallinity does not always guarantee high thermal conductivity if the crystallites are not well aligned in the polymer.[92, 123] While the crystalline domains should have much higher thermal conductivity than their amorphous counterparts,[112] the existence of amorphous phases between these crystalline domains can present large thermal resistance.[7, 110] In addition, the crystal-amorphous interface can also scatter phonons.[112] It is not clear how large a role such interfacial thermal resistance plays in the overall thermal conductivity. It has been shown that if the amorphous region consists of linker molecules, the crystal-amorphous interface is connected by covalent bonds, which present much smaller resistance than the amorphous region itself.[64] However, if the interface is connected by purely weak vdW forces, the relative importance of the interfacial resistance will depend on the size of the amorphous domain, which might still dominate the overall thermal resistance given its low thermal conductivity.

It should be pointed out that amorphous phases in polymer fibers can have a certain degree of chain alignment or orientation preference due to its synthesis process, which is different from the fully random amorphous structure in bulk polymers. External forces like stretching can align the chains in amorphous regions and give them higher order, and large forces will straighten the chains in the amorphous region, which will facilitate thermal transport.[110] Robbins et al. measured the MFP of polymer fibers with transient grating (TG) spectroscopy.[126] They pointed out that domain boundary scattering instead of phonon-phonon and phonon-defect scattering contributes most to the thermal resistance in polymer fibers. Moreover, amorphous regions can allow the travel of low-energy phonons, which help the propagation of thermal phonons between crystalline areas. Consequently, many phonons can travel across domain boundaries. An MFP up to 200 nm was observed for a semicrystalline PE fiber with a draw ratio of 7.5.[126]

### 2.7.3. IMPACT OF INTER AND INTRA CHAIN INTERACTIONS

The types of inter- and intra-chain bonding (vdW or covalent bonds) and their relative ratio in the fiber can influence the thermal conductivity of polymer fibers. The contribution towards thermal conductivity from covalent bonds is much higher than that from weak forces like vdW forces.[76] Stronger inter-chain vdW and large dihedral angle energy will confine the chains and limit the rotation of chain segments,[52, 56] giving rise to higher thermal stability and thermal conductivity.[56] However, inter-chain vdW interaction can also scatter phonons inside the chain, impeding thermal transport.[127] Apparently, the above-mentioned competing effects coexist. Intra-chain vdW forces would mostly work adversely to thermal transport as it can lead chains to coil, impairing the thermal transport efficiency of the covalent backbone but itself has limited contribution to thermal conductivity.[128] Such effect is better demonstrated in amorphous polymers as will be discussed in Section 3.5.3.

### 2.7.4. IMPACT OF CHAIN CONFORMATION IN POLYMER FIBER

Chain conformation, which can be influenced by temperature or the inherent chemistry of molecules, is the root cause of the displayed thermal transport properties of polymer fibers. Even a subtle change in polymer chain conformation can lead to a significant change in thermal conductivity for highly aligned polymer fibers.[12, 52, 56, 66] Below, we discuss some important factors influencing chain conformation.

**Sidechains:** For achieving a high thermal conductivity, it is desirable to use long chains without or with few short side chains.[59, 129] Side chains can lower thermal conductivity as they can serve as scattering centers for phonons transport along the backbone. Side chains connected to the backbone will present a different bonding environment to the bonding atom than the rest on the backbone, which leads to defect scattering. In this sense, it was found that lighter and more symmetric side groups are less detrimental to phonon transport.[125] Simulations on bottlebrush polymers showed that longer side chains led to more disorders in the polymer and reduced thermal conductivity.[130] Thermal conductivity of bottlebrush polymers with aligned backbones decreases sharply with the increasing side chain length and eventually converge to a low value, which can be explained by the fact that the interchain scattering among those long side chains increases.[130] It was shown with the XRD pattern that longer side chains led to an overall polymer morphology between crystalline and amorphous phase, where an amorphous-like structure appeared in those excellently aligned polymer bottlebrush fibers.

**Backbone:** When considering the polymer chain backbones, stiffer ones usually lead to higher thermal conductivity.[16] As a result, it is shown that high modulus polymer fibers generally have higher thermal conductivity. The high moduli of polymer fibers originate from the strong covalent bonds along the backbone. Such strong intra-chain bonding can lead to higher phonon group velocity. However, there are also exceptions. For example, Kevlar, which has higher modulus than PE, displayed much lower thermal conductivity than PE. It was found that in Kevlar the energy of one of the dihedral angles along the backbone is exceptionally weak, and segmental rotation happens frequently around this dihedral angle, leading to disorder scattering.[56] Other MD simulations of polymers (e.g., PDMS and PE) also revealed the same mechanism.[52, 131] Another MD work predicted that Kevlar fibers, if fully stretched to eliminate the segmental rotation, the thermal conductivity could be increased from 11.01 W/mK to 147.99 W/mK.[55] It was also found that large chain segmental rotation can reduce the phonon group velocity besides phonon MFP, which also contribute to thermal conductivity reduction (**Fig. 8**).[55] Chemical structures like double bonds or $\pi$-$\pi$ stacking interactions are highly useful for improving the polymer chain's stiffness and reducing segment rotation, which offered useful insight to the structure-property relation for designing thermally conductive polymer fibers.[56, 59, 104] In addition, thermal conductivity can also be improved by restricting the angular bending freedom.[132]

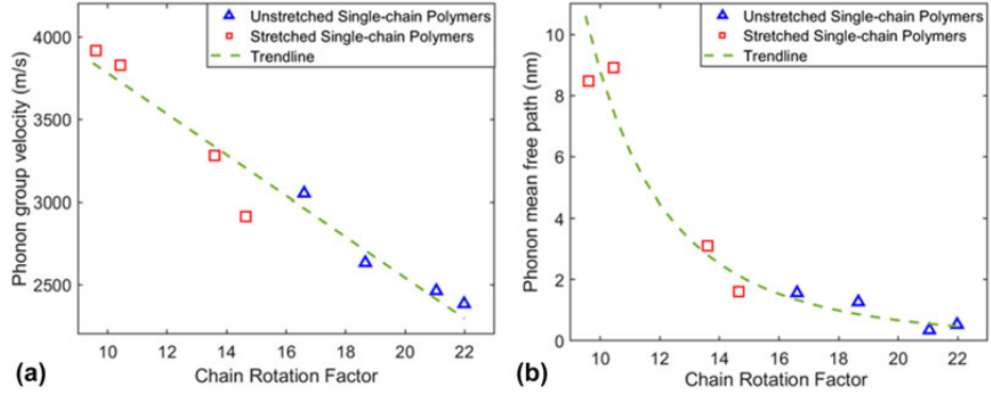

**Figure 8.** Phonon group velocity and MFP as a function of chain rotation factor (CRF), defined as $CRF = \frac{1}{N}\sum_{i}^{N}\frac{1}{PDE(y,z)}$, where $i$ denotes the $i^{th}$ atomic position; $N$ means atoms in a given single polymer chain along the x-axis; $PDE(y,z)$ denotes the probability density estimate for each atom $(y,z)$ based on Kernel density estimator.[55] Reproduced with permission.[55] Copyright 2018, Cambridge University Press.

**Chain confinement:** Spatial confinement can influence chain conformation and thus thermal conductivity in polymer fibers. This is why thinner fibers usually have larger thermal conductivity than thicker ones.[88] For thinner polymer fibers, polymer chains are forced to orient in the longitudinal direction, which favors phonon transport along the strong backbone.[86] For thicker polymer fibers, chains have more room to randomly orient, and thus phonons transport will be more isotropic, which lowers the thermal transport efficiency in the fiber direction. Note that the effect of chain confinement has a different effect on polymer fibers vs. amorphous polymers. In ultrathin amorphous polymer thin films, the thermal conductivity increases with $\frac{dz}{Rg}$, where dz is film thickness and Rg is the radius of gyration. In other words, stronger confinement leads to lower thermal conductivity in amorphous polymers.[133] TDTR measurement of PS ultrathin films showed that in the confined region, the thermal interfacial conductance is enhanced because of less entangled chains near the PS/substrate interface.[134]

### 2.7.5. IMPACT OF MOLECULAR WEIGHT

Molecular weight is determined by the degree of polymerization and the mass of the constituent atoms. A higher degree of polymerization leads to longer polymer chains, which usually results in higher thermal conductivity.[39, 59, 95, 125] For short chains, phonon transport along the strong backbone can be mostly ballistic and mainly get scattered at the chain-ends for short chains (e.g., PE with chain segments N<50).[51, 135] For longer polymer chains (e.g., PE with N>200), the dominant mechanism can instead be from factors like phonon-phonon scattering and structural disorder (e.g., segmental rotation) scattering. This was experimentally illustrated by using γ-ray to cut the polymer chains shorter in polymer fibers.[136, 137] The reduction in the degree of polymerization in a UHMWPE fiber from 1700 to 200 halved its thermal conductivity. Besides, the presence of heavy atoms in the monomer can lead to lower thermal conductivity since they will lead to lower phonon group velocity.[56] A parametric MD study indicated a negative relationship between atomic mass and thermal conductivity of a model polymer.[67]

### 2.7.6. DEFECTS AND IMPERFECTIONS

The main reason for polymer fibers to have much higher thermal conductivity than their bulk counterparts is the reduction of defects and imperfections.[57, 59, 129, 131, 138] From a phonon transport point of view, defects in the polymer can include molecular level defects like chain end, kinks, entanglements and random orientations, and extrinsic defects like voids, boundaries, dislocations, and amorphous-crystalline interfaces. All these defects serve as phonon scattering centers. Like drawing, the fabrication process can largely eliminate large extrinsic defects like voids as chains become more aligned. The molecular level defects can scatter phonons significantly in the absence of extrinsic defects. For example, it was found that kinks in the backbone can twist polymer chains and reduce thermal conductivity dramatically (**Fig. 9**).[131] MD simulations prediction showed that a perfect PE chain with no defects could have thermal conductivity as high as 1400 W/mK and that of a defect-free PE fiber could exceed 300 W/mK.[57, 129] However, the highest thermal conductivity experimentally obtained is much lower (~104 W/mK) even ultra-drawing eliminated most of the extrinsic defects.[88]

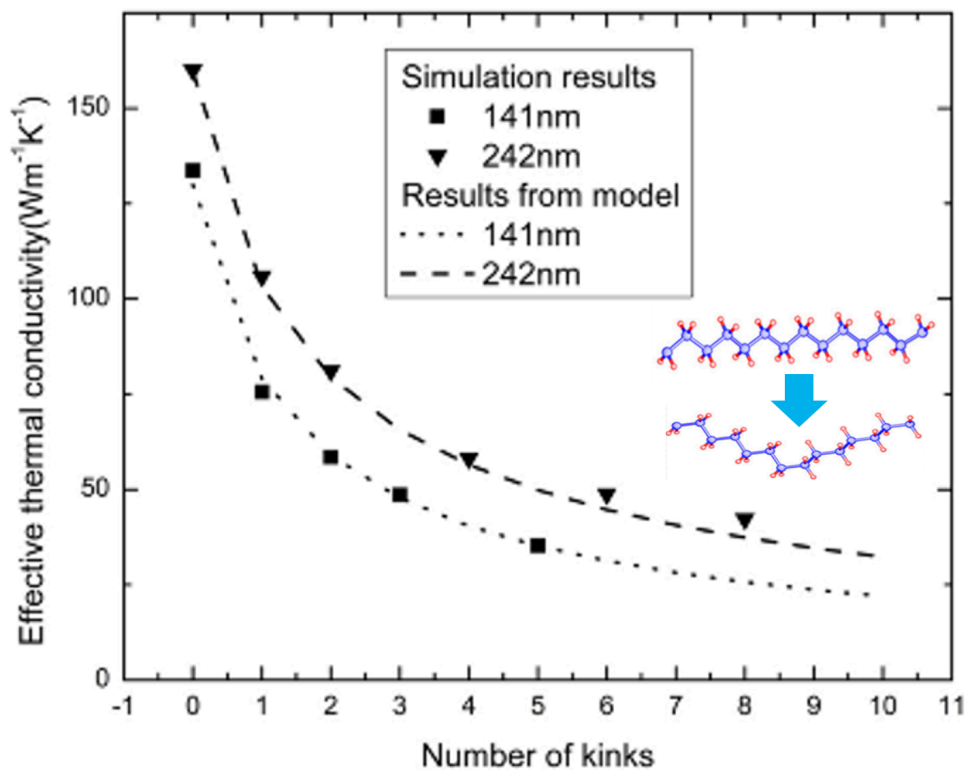

**Figure 9.** Effect of number of kinks on thermal conductivity of a polymer chain.[131] Reproduced with permission.[131] Copyright 2019, AIP Publishing.

## 3. BULK POLYMERS

It is well-known that bulk amorphous polymers are not good thermal conductors with the thermal conductivity mostly in the range of 0.1-0.5 W/mK at room temperature.[139] The low thermal conductivity has made polymers ideal candidates for thermal insulation,[140, 141] but in many other applications, higher polymer thermal conductivity is desirable. Industry mostly uses polymer composites for heat transfer-critical applications, such as thermal interface materials and plastic heat exchangers. In many applications, polymer thermal conductivity greater than 10 W/mK is desired, but realizing this value has been proven difficult without compromising other properties of polymers (e.g., electrical insulation). While the thermal conductivity of inorganic fillers and their morphology in the composites, such as dispersion and percolation,[142] influence the overall heat transfer performance of the composites, the thermal conductivity of the polymer matrix can be a limiting factor in improving the overall composite thermal conductivity. For example, if we consider compositing high thermal conductivity particles with a polymer matrix (**Fig. 10**), effective medium theory calculation[143, 144] easily shows that there will be a significant difference in the composite thermal conductivity whether the polymer matrix has low (e.g., 0.15 W/mK) or high (1.5 W/mK) thermal conductivity. While of course this simple model ignored factors like interfacial resistance, percolation and anisotropy, it does underline the importance of the thermal conductivity of the base polymers.

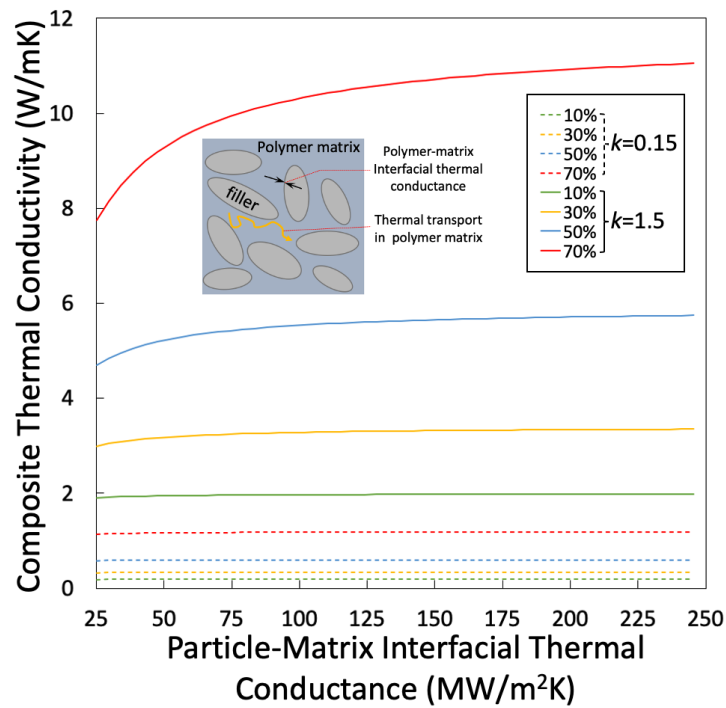

**Figure 10**. Effective medium theory calculation of polymer composite thermal conductivity with BN fillers (assumed to be isotropic) as a function of interfacial thermal conductance between the filler and the matrix. Polymer matrices with thermal conductivity of 0.15 W/mK and 1.5 W/mK are considered. Different color lines correspond to different volumetric loading fractions. It is seen that improving composite thermal

Thermal Transport in Polymers: A Review. 31

conductivity beyond 10 W/mK requires the matrix thermal conductivity to be 1.5 W/mK. Interfacial thermal conductance only starts to matter when the matrix thermal conductivity is high and filler fraction is high.

It is also worth mentioning that thermal conductivity of composites is rarely a standalone factor to consider for thermal applications. For example, in the electronics industry, chip manufacturers always want to avoid electrically conductive fillers in thermal interface materials due to the risk of short-circuiting, leaving compositing polymer with metallic fillers (e.g., silver particles) not always viable, but metal-polymer composites usually have higher thermal conductivity than ceramic-based ones. Carbon nano-materials (e.g., carbon nanotubes and graphene) are promising fillers,[33] but they tend to be much more expensive and their dispersion is not as well controlled as conventional fillers. In addition, since the primary purpose of thermal interface materials is to fill the air gaps between two rough surfaces, the polymer composite needs to be soft so that it can conform to the surface landscape, but softer polymers usually have lower speed of sound and lower thermal conductivity than stiffer ones.[145, 146] Moreover, loading excessive amount of fillers also stiffens polymers, impairing their ability to conform to surfaces. As a result, designing polymers for heat transfer applications is a highly constrained design task.

Nevertheless, even for thermal conductivity of amorphous polymers alone, design principles are lacking due to the lack of understanding of the thermal transport physics in polymers at different levels. In this section, we review the historic and current macroscopic, microscopic and molecular-level understanding of polymer thermal transport. For the macroscopic understanding, observing relations between polymer thermal conductivity and other properties are the main route, but such studies can be highly phenomenological. Microscopic understanding emphasizes more thermal transport physics from the heat carrier point of view, enabling some predictive power of models derived from these studies. Finally, we discuss the recent advancements from the molecular-level, which sheds light on the structure-property relationship for polymer thermal conductivity from experiments and molecular simulations. We believe the molecular-level studies are key to understanding and predicting thermal conductivity of polymers, and eventually to achieving the overarching goal of materials by design.

### 3.1. TEMPERATURE DEPENDENCY

In early studies, correlating measured thermal conductivity of polymers with their other properties has provided important insights to thermal transport physics and guidance to materials engineering. These other properties mainly include temperature, density, specific heat, pressure, crystallinity, glass transition temperature, and etc.[147-152]

Among them, the thermal conductivity dependence on temperature has been the most important source for the understanding of microscopic heat transfer physics in polymers. The understanding of polymer thermal conductivity largely borrowed the physical picture for amorphous inorganic materials. The explanation of thermal conductivity in amorphous dielectrics started from Kittel,[153] who used a variant of the solution to the phonon Boltzmann transport equation, $\kappa = cvl/3$,[153] where $c$ is volumetric heat capacity, $v$ the average phonon group velocity and $l$ the MFP. This is essentially the same formula for describing any phonon gas thermal conductivity, but the difference lies in $l$. By analyzing a number of amorphous glasses, $l$ was determined to be approximately a constant of ~7 Å. The physical interpretation is that amorphous materials have random networks of atoms, and when the dominant phonons have wavelengths shorter than the characteristic length of the atomic network, boundary scattering of the unit cells would lead to a constant $l$ roughly equal to the atomic scale characteristic length.

However, a constant MFP cannot explain the thermal conductivity trend at the low temperature limit (0.1-1 K). It has been observed that thermal conductivity of all amorphous polymers has similar magnitude and a characteristic trend of $\sim T^2$ below 1 K (**Fig. 11a**), similar to those observed in amorphous inorganics in the same temperature range ($\sim T^{1.8}$).[154] At low temperatures, heat capacity $c \sim T^3$, while the excited phonons have the same $v$, and thus the temperature dependency of $l$ would determine the overall behavior of $\kappa$. When the temperature is sufficiently low, the dominant heat carriers are long wavelength phonons, much longer than the atomic characteristic lengths. In such situations, amorphous materials behave like elastic media, where the detailed atomic structure is not important to the transport of these long wavelength phonons. The argument was that these modes thus would have long MFPs since they cannot be effectively scattered by the structure and the anharmonicity is insignificant at low temperatures. However, this would lead to no difference between the thermal conductivity behaviors of amorphous materials and crystalline lattice, whose thermal conductivity scales with $T^3$ at low temperatures, where $l$ is bounded by boundary scatterings (**Fig. 11a**).

Klemens, in one of his early seminal works,[155] introduced the idea of structure scattering due to the disorder of the amorphous structure. He argued that while disordered structures lack regularity, instantaneous displacement of atoms can still be projected into plane waves, but these plane waves interact with each other inelastically even at the harmonic limit. It is noted that in crystals, these plane waves do not interact with each other at low temperature where anharmonicity is weak. He further argued that for modes with wavelength much larger than the lattice constant, phonons behave like propagating modes in crystals, but for modes with wavelength smaller than the lattice constant, the MFP is a constant. This model led to a $\kappa \sim T^1$ relation at the low temperature limit, while $\kappa \sim c$ (heat capacity) at high temperatures. Ziman[156] had proposed a similar model. Both models showed that for long wavelength phonon MFP, $l \sim \omega^{-2}$, and since $c \sim T^3$ and $v$ is constant, using Kittel's model ($\kappa = cvl/3$) will eventually give the $\kappa \sim T^1$ (note: this conclusion used the fact that the dominant phonon to thermal transport has $\omega \sim T$ considering a linear dispersion[157]). Despite the discrepancy between Klemens' theory and the experimental trend observed at extremely low temperatures ($\sim T^2 < 1$ K[154]), his model managed to capture thermal conductivity of quartz glass in a wide temperature range of $\sim 4 - 90$ K.[155] It is noted that the experimental data used in his work did not extend below 1 K.

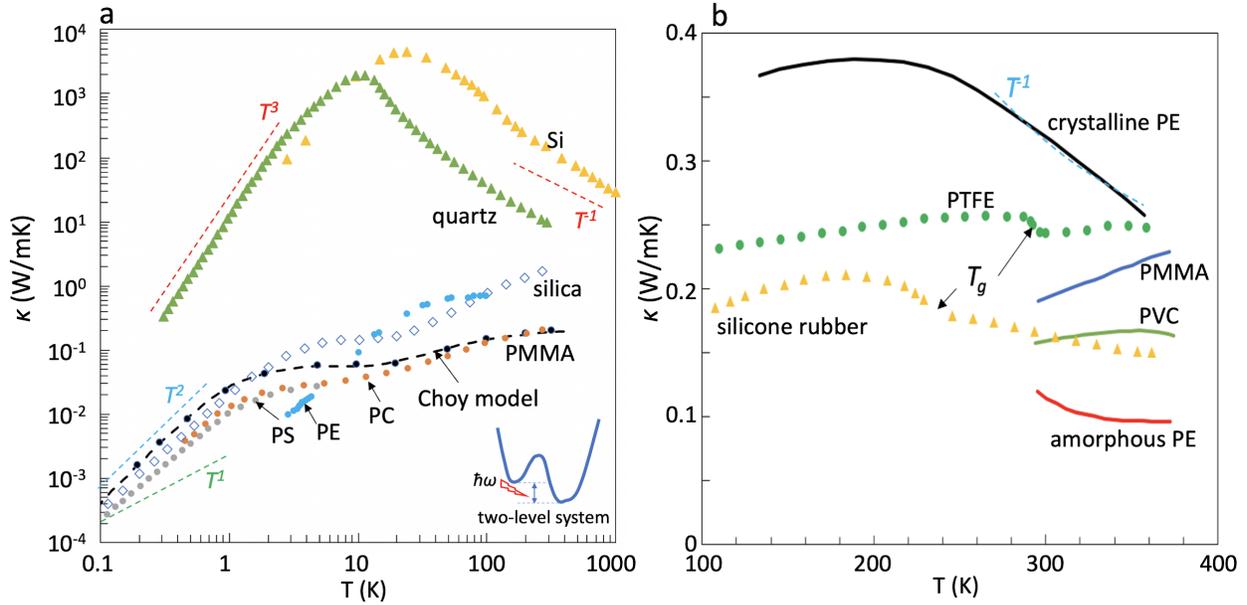

**Figure 11. (a)** Thermal conductivity of some common amorphous polymers (PMMA (polymethyl methacrylate), PS (polystyrene), PC,[158-160] and PE[161] as a function of temperature with the focus on low temperatures. The Choy model is shown as a black dashed line for PMMA.[152] At very low temperatures (< 1 K), the amorphous polymer thermal conductivity scales with $T^2$. As a comparison, amorphous inorganic materials, silica, has a similar temperature dependency as amorphous polymers. Crystalline materials, however, show very different temperature dependency (~$T^3$ at low temperature) compared to amorphous materials, since their thermal carrier scattering mechanisms are different. The schematic inset shows a two-level system Anderson[162] and Phillips[163] used to explain the $T^2$ dependence of thermal conductivity at extremely low temperatures (< 1K). **(b)** Thermal conductivity of common polymers, including amorphous PE, PMMA, PVC,[151] crystalline PE,[150] and PTFE and silicone rubber across the glass transition temperature.[150] Polymers can exhibit distinct temperature dependence of thermal conductivity at high temperature depending on their crystallinity, glassy state and etc.

To explain the low-temperature $\kappa \sim T^2$ trend, Anderson et al.[162] and Phillips[163] proposed a resonant scattering mechanism in a two-level system. They assumed that certain atoms in the amorphous structure experience double-minima potential landscapes and the transition from one minimum to another depends on the interaction between this two-level system and low frequency phonons (inset in **Fig. 11a**). When a phonon impacting the atom in the two-level system has a large enough energy to help the atom overcome the energy barrier, the phonon energy is absorbed and incoherently re-emitted, which leads to an inelastic scattering of the phonon. They have derived that the phonon MFP due to such scattering follows $l \sim \omega^{-1}$. When considering that the dominant phonon frequency is proportional to temperature at low temperatures (i.e., $\omega \sim T$), one can derive $l \sim T^{-1}$, and then the model would successfully yield $\kappa \sim \frac{cvl}{3} \sim T^3 T^{-1} = T^2$ (**Fig. 11a**). Combining Anderson's model for long wavelength modes and constant MFP for higher frequency modes, Choy[152] successfully fit the whole curve for PMMA thermal conductivity from 0.1 to 300 K. The interesting plateaus universally appear for amorphous materials, including both organics and inorganics (**Fig. 11a**), around 5-10 K turns out to be a simple superposition effect of the long and short wavelength phonon contributions to the thermal conductivity. It is worth mentioning that from fitting experimental data of polymers, Choy[152] found the constant MFP to be around 7.2 Å for PMMA, which is very similar to that of the inorganic glass as found by Kittel.[153]

At intermediate temperatures (~10-200 K), where wavelength of the dominant phonon is shorter than the characteristic length of microscopic disorder, $l$ is a constant according to Klemens[155] and Kittle,[153] and thus thermal conductivity behaves the same as heat capacity, which increases slowly with temperature at ~10-200 K.

At even higher temperatures (> 200 K), however, the temperature dependence of thermal conductivity varies significantly from one polymer to another even for the most common ones (**Fig. 11b**). Some polymers have increasing thermal conductivity as temperature increases (e.g., PMMA, polyvinyl chloride (PVC)),[151] but some other polymers (e.g., amorphous PE[151] and silicone rubber[150]) have decreasing thermal conductivity as a function of temperature (**Fig. 11b**). Even for the same type of polymer (e.g., PE), with different morphologies (e.g., crystalline vs. amorphous PE), the temperature trends differ. It can be argued that for crystalline polymers, the thermal conductivity is similar to crystalline inorganic lattices and the thermal conductivity should decrease due to enhanced anharmonic phonon scattering. As shown in **Fig. 11b**, the high temperature behavior of crystalline PE can be roughly described by $T^{-1}$ – the same as those of silicon and quartz also shown in the figure. For amorphous polymers, however, the reason for their temperature trend has been attributed to the decrease in density due to thermal expansion above the glass transition temperature, $T_g$. Hattori[164] proposed that above the glass transition temperature, micro-Brownian motion of polymer molecules became excited, which led to the increase in the molecular mobility, but we are not clear why such enhanced mobility decreases thermal conductivity. The increased mobility should have increased the thermal transport contributed by advection. However, recent simulations have found that such advection terms are negligibly small in thermal conductivity, and the decrease in thermal conductivity is associated with the decrease in density as temperature increases.[111] In the book from van Krevelan and Nijgenhuis,[165] thermal conductivity values of ten polymers from very low temperature to temperature higher than $T_g$ was plotted, and an phenomenological model from Bicerano[166] was able to fit the data trend for the whole temperature range well. More recently, an empirical model was developed for thermal conductivity of amorphous polymers based on the kinetic theory, where density, monomer molecular weight, and sound speed were the only parameters needed for predictions.[167] The model agreed well with experimental data when it comes to temperature-dependence of thermal conductivity.

### 3.2. DENSITY AND HEAT CAPACITY EFFECT

It has long been observed that polymer thermal conductivity increases with density. As a first approximation, Hands et al.[151] used the thermal conductivity model for liquids[168] $\kappa \sim \rho^{\frac{4}{3}}$ for polymers above the glass transition temperature, where $\rho$ is density. By surveying a number of polymers with different densities, it seems that such a relation can reasonably describe the thermal conductivity trend as a function of density (**Fig. 12a**). This model would predict a decreasing trend as a function of temperature. Such a decreasing trend is fundamentally different from that in lattice thermal conductivity due to anharmonic phonon scattering.

However, if we fit the data using a power law, it would yield $\kappa \sim \rho^{0.9}$ (**Fig. 12a**). We would like to cast our doubts on the application of the model for liquid thermal conductivity to polymers. We have recently found that thermal transport along the polymer chain backbone contributes more significantly to thermal conductivity via the strong intra-molecular covalent bonding interactions than inter-molecular interactions (**Fig. 12b**).[111] In contrast, thermal transport in liquids is due to advection and inter-molecular energy transfer. It has been proven that the change in density (**Fig. 12c**) is accompanied by the change in radius of gyration ($R_g$) of polymer chains, which influences the thermal transport along the polymer chain backbone (**Fig. 12d**). When density decreases, the chains have more room to move, which can influence the $R_g$ due to the competition of enthalpic and entropic effects,[111] and thus change the contribution from the intra-chain interaction to thermal conductivity. In the meantime, the reduced density enlarges inter-molecular distance which will lead to a decrease in thermal conductivity contributed by the inter-molecular interactions.[111] These two competing effects would lead to a trend slower than $\rho^{\frac{4}{3}}$ (e.g., $\rho^{0.9}$ seen in **Fig. 12a**) which was derived for liquids. It is clear from the MD simulation results in **Fig. 12c-d**, thermal conductivity follows a trend that resembles that of $R_g$ as a function of temperature. Of course, $R_g$ also depends on the type of polymers, and thus might not be fair to simply say it would be larger for lower density polymers. However, it is our belief that the thermal conductivity dependence on density of polymers is much more complicated than that in simple liquids. In addition, the impact of density on the inter-molecular thermal transport alone would not be universal since different types of interactions (e.g., vdW and Coulombic interactions) have different decay rates as the interatomic separation enlarges (e.g., vdW ~ $r^{-6}$, and Coulombic ~ $r^{-1}$), which are found to be critically important in interpreting the trend of polymer thermal conductivity.[169]

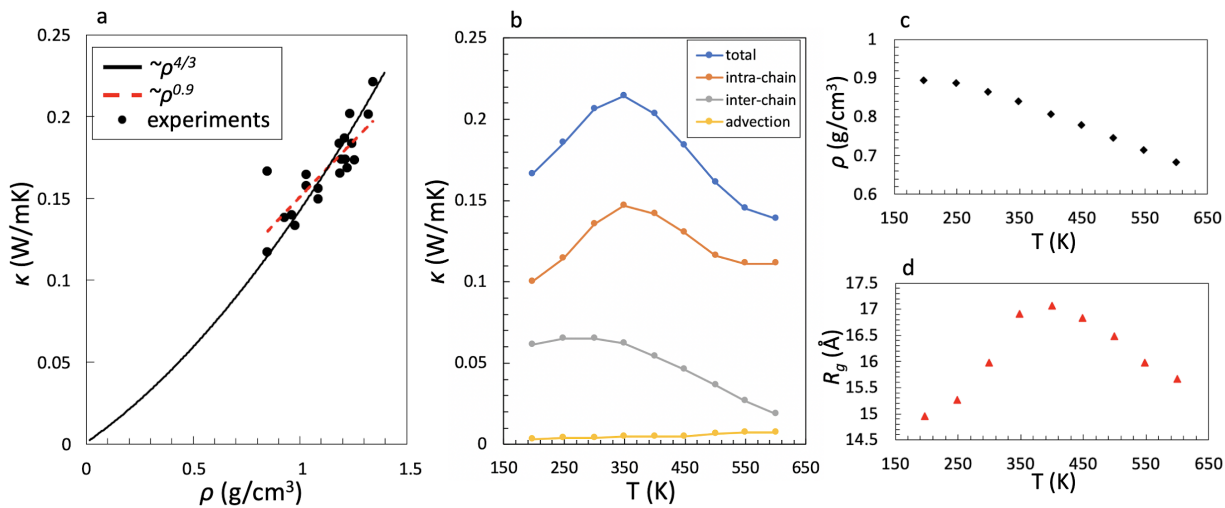

**Figure 12. (a)** Thermal conductivity of different polymers as a function of their densities.[151] **(b)** Thermal conductivity as a function of temperature decomposed into inter-molecular, intra-molecular and advection contributions from MD simulations of amorphous PE.[111] **(c)** Density and **(d)** radius of gyration ($R_g$) as a function of temperature from MD simulations.[111]

Besides temperature, external pressure can also change polymer density. By fitting the experimental data, it is found that the pressure effect on thermal conductivity of amorphous PMMA was related to the elastic constant and the atom number density of the polymer.[170] Density has been linked to thermal conductivity of solid polymers via its impact on volumetric heat capacity. In the classical limit, there are $3N$ ($N$ is the number of atoms) vibrational modes in a system and higher density leads to higher volumetric heat capacity. However, vibrational modes in reality are not equally excited due to quantum effect. Olson et al.,[171] Wang et al.[172] and Xie et al.[145] argued that since hydrogen vibration are not excited at room temperature according to the Bose-Einstein distribution, atom number density used towards predicting thermal conductivity should exclude hydrogen atoms, and after such a treatment Xie et al.[145] indeed managed to improve the agreement of the minimum thermal conductivity model[173] and experimental measurements for 18 polymers, but the model results are still always larger than measurements. The authors had to include corrections concerning the localized modes to bring better agreement between the model and measurements.[145] We would like to note that these treatments may need further study. For example, even carbon vibration modes are not all excited at room temperature, and categorizing modes associated with atoms involved in loop structures as localized modes might not be physical since these atoms still interact with other atoms and can transfer energy via the inter-atomic interactions. A recent mode-resolved study showed that contribution from localized modes in disordered $SiO_2$ is non-negligible due to their correlation with the spatially extended but non-propagating modes (diffusons),[174] and we expect non-negligible contribution from the localized modes in polymers as well.

We note that recently Xi et al.[175] proposed a universal formula to calculate the thermal conductivity of materials ranging from crystals to amorphous polymers. For amorphous polymers, the formula leverages a network theory developed by Zhou et al.[176] In their physical picture, the amorphous polymer system is treated as a percolation network of segments of polymer chains. Heat conduction depends on the density and interaction strength of the connection points in the network. It was assumed that the heat transfer efficiency is determined by the van der Waals (vdW) interaction between the segments connected to each connection point. The model could explain well the universal low thermal conductivity in amorphous polymers and its temperature and pressure dependence. However, questions remain on why the segments connected to each point have the same heat transfer efficiency that is dictated by vdW interactions, while at least two of the four segments are connected by covalent bonds. As discussed previously, molecular simulations have shown that thermal transport along the covalent backbone usually dominates even in the amorphous phase. It is possible that if such local bonding anisotropy can be implemented, the model may reach higher quantitative accuracy. Alternatively, maybe some effective connection point density should be used other than the actual physical connection point density since polymer chains in the molecular level are almost closely packed.

For expanded polymers which become porous, density effect on thermal conductivity is governed by the void fraction and can be reasonably described by effective medium theory considering a composite of air and polymer.[151]

### 3.3. CRYSTALLIZATION EFFECT

Many polymers are semi-crystalline and this could have effects on the apparent MFP, $l$, since heat carriers traveling in the crystal domains would be subject to less structural disorder scattering and thus have longer $l$, which will lead to higher thermal conductivity. Hattori[177] found a linear relationship between thermal conductivity of polytrifluorochloroethylene (PCTFE) and the degree of crystallinity at room temperature (**Fig. 13b**). He made an argument that when there were more crystalline domains in the polymer, the effective MFP of heat carriers were larger and thus the thermal conductivity increased when considering the phonon gas model (i.e., $\kappa = cvl/3$). Crystallization can happen spontaneously given proper heat treatment (e.g., annealing). In such cases, the thermal conductivity would still be isotropic since the crystallites orients randomly. As discussed previously in Section 2.4, crystallites orientation can be forced via extrusion or drawing, which induces shear to the internal structure of polymers (**Fig. 13a**). Even with small draw ratios, experimental characterization has shown that crystallinity can be high and crystalline lamellae will start to align well in the draw direction.[178, 179] Further drawing will stretch the tie molecules between the lamellae and pull out crystalline blocks. When the drawing ratio further increases, these crystalline blocks will further align along the draw direction.[152]

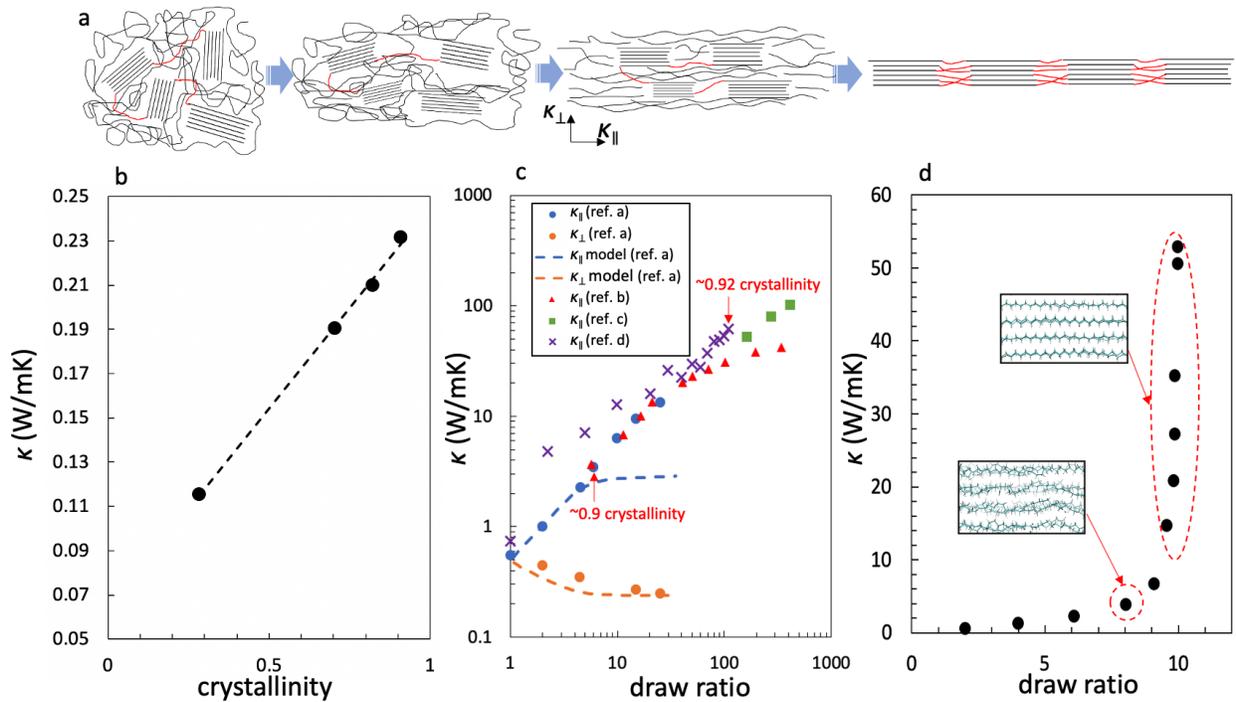

**Figure 13. (a)** Schematic of the internal structure changes when the polymer is subject to drawing. **(b)** Thermal conductivity as a function of crystallinity for PCTFE.[177] **(c)** Thermal conductivity along the draw direction ($\kappa_\parallel$) and perpendicular to the draw direction ($\kappa_\perp$) at different draw ratios. Refs.: a:[180]; ref. b:[38]; ref. c:[4]; ref. d:[110]. **(d)** Thermal conductivity of PE at different draw ratios from MD simulations. Sharp change in thermal conductivity at a narrow draw ratio window is due to the segmental ordering along the aligned chains.[66]

Choy and Young[181] developed an effective medium formula for fitting thermal conductivity of semi-crystalline polymers by considering crystalline domains embedded in an amorphous matrix as a two-phase composite. The model agrees favorably with experimental data for poly(ethylene terephthalate) (PET) and PE at crystallinity up to ~75%, but starts to underestimate the thermal conductivity afterward. The thermal conductivity of the crystalline domain was left as a fitting parameter. The model assumes that the matrix phase has an isotropic thermal conductivity that is the same as pure amorphous polymer, and the discrepancy at high crystallinity samples was attributed to the fact that molecules starts to form inter-crystalline bridges from the tie molecules (red lines in **Fig. 13a**). Choy et al.[180] further related the anisotropic thermal conductivity of different polymers to their draw ratios to infer how crystallinity and crystal orientation influence thermal conductivity. They used their modified Maxwell's effective medium approximation to fit the anisotropic thermal conductivity. While the tie molecules were hypothesized to influence the overall thermal conductivity, the underlying mechanism only recently starts to be revealed.[64]

As shown in **Fig. 13c**, it is interesting to find out that even with low draw ratio (~5), the crystallinity of the polymer is already 0.9, where the thermal conductivity is around 2 W/mK.[38] Further increasing the draw ratio to over 100 in thin films only increase the crystallinity slightly, but the thermal conductivity increase is tremendous to ~ 60 W/mK.[110] Ultra-drawing fibers into nanofibers was shown to further increase the thermal conductivity to ~100 W/mK.[4] However, the room for crystallinity increase is small after certain draw ratios (~5), and it has been found that the crystallite orientation along the draw direction also saturates after relatively low draw ratio of ~10,[110] after which thermal conductivity continued to increase by 10 folds (**Fig. 13c**, purple crosses). SAXS (Small-angle X-ray scattering) and WAXS (Wide-angle X-ray scattering) analyses indicated that increasing the draw ratio beyond ~10 will lead to the decrease in the amorphous volume fraction, which effectively increases crystallinity, but such increase is not significant. Thus, this could still not explain the high thermal conductivity increase. Using a serial thermal resistance model, the authors concluded that the thermal conductivity of the amorphous portion is increasing when the draw ratio increases, and at a draw ratio of ~110, the amorphous thermal conductivity is calculated to be around 15 W/mK.[110] It was noted that the amorphous region is not crystallized but chains are somewhat orientated. Lu et al.,[64] using MD simulations to study semi-crystalline PE, found the thermal conductivity of the amorphous region needed to be modified due to the bridge (tie) molecules between the two neighboring crystallites in order to properly describe the semi-crystalline PE thermal conductivity especially at high crystallinities (>83%). The thermal conductivity of semi-crystalline PE depends on the number of tie molecules bridging the crystallites. It might also be possible that after the tie molecules are stretched, their chain segments along the backbone become more ordered, which is critical to enhancing thermal transport along polymer chains as revealed by another MD simulation,[66] which showed that extended polymer chains can display a step-like thermal conductivity after a certain draw ratio (**Fig. 13d**).

### 3.4. SPEED OF SOUND EFFECT

Speed of sound has been another important factor that has been commonly linked to polymer thermal conductivity (i.e., $\kappa = cvl/3$), besides heat capacity and MFP. Sound speed is related to the modulus of materials and thus we usually see harder materials having higher thermal conductivity. For example, diamond and cubic BN, the hardest materials, are known to have among the highest thermal conductivity in nature.[182, 183] Modulus can be measured from conventional tensile mechanical test, which can be used to calculate speed of sound. Speed of sound can also be measured directly using acoustic echoes in pump-probe measurements where thermal conductivity can be characterized in the same experiment.[146] **Figure 14** shows some experimental and MD thermal conductivity data for polymers as a function of the average speed of sound. Here, the average speed of sound is calculated as $V_{ave} = 1/3(V_l+2V_t)$, where $V_l$ and $V_t$ are respectively the speeds of the longitudinal and transverse acoustic modes (i.e., phonon group velocities at Brillouin Zone center). As can be seen, there is a generally increasing trend of thermal conductivity against the average speed of sound. Data shown in **Fig. 14** are also tabulated in **Table 3.**

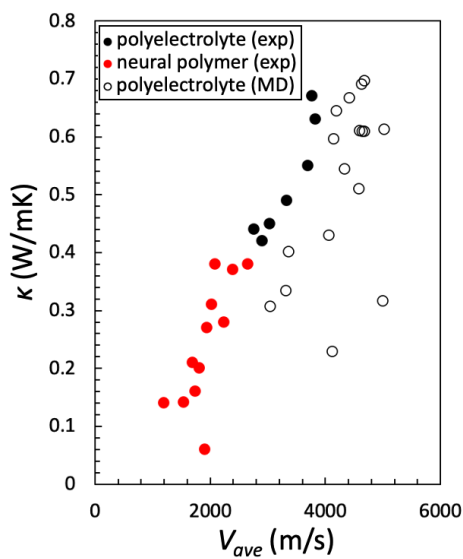

**Figure 14**. Thermal conductivity of polymers as a function speed of sound. MD data are from Ref. [184], and experimental data are from Refs. [145, 146]. Data shown in the figure are also tabulated in **Table 3.**

However, there are doubts on the accuracy of using speed of sound to describe the heat carrier traveling speed. Heat carriers in disordered structures have a wide frequency spectrum and only those with very low frequencies (i.e., long wavelength) can be reasonably assigned with the speed of sound as their traveling speed. For those intermediate frequency and higher frequency modes, they should have lower traveling speed. It might be reasonable to use $\kappa = cvl/3$ or its variations for very low temperature thermal conductivity when long wavelength phonons dominate. At higher temperatures, the dominant frequency shifts to a higher region, which should have lower traveling speed compared to low frequency modes. It was pointed out that only at very low temperature (<1 K), heat capacity approaches the value predicted on the basis of the sound velocity.[185] Allen and Feldman[186] categorized the heat carriers in amorphous materials into three kinds depending on their nature: *propagons* (phonon-like propagating wave delocalized over large distances), *diffusons* (extended vibration modes delocalized over a short distances) and *locons* (highly localized modes). Depending on temperature, morphology and materials, these modes can contribute differently to heat conduction. In terms of population, the diffusons and locons are the majority. According to Allen and Feldman's calculation, diffusons, rather than the long wavelength propagons, are by far the most dominant heat carriers at room temperatures.[186] Diffusons transport fundamentally differently from propagons since they transfer energy via diffusion-like mechanism between extended but non-propagating modes, a picture that is different from the phonon gas model. Alexander et al. described transport of such non-propagating modes as anharmonic coupling.[187] Lv et al.[174] showed that diffusons can also correlate to locons so that the energy of locons can be effectively transferred spatially with the help of diffusons. It is hard to rationalize that these diffusons or locons would travel with speed of sound, and their transport might be better described by a correlation picture.[188] However, another recent study showed that propagons were the dominant contributors to thermal conductivity in amorphous silicon,[189] and these modes are mainly limited by scattering from local fluctuation of elastic modulus rather than anharmonicity – a picture similar to that proposed by Klemens, who considered spatial variation of speed of sound.[190, 191] It is not clear if the dominant thermal carrier natures will be different in amorphous polymers compared to amorphous silicon, but a study comparing Allen-Feldman's model to MD simulations on PS implied so.[192] Another study on amorphous carbon, which is closer in elemental composition to polymer than silicon, showed that propagons contribute virtually nothing to the thermal conductivity at room temperature.[193]

**Table 3**. Summarized thermal conductivity and speed of sound for different polymers. Red colors are neural polymers and black colors are polyelectrolytes. MD data are from Ref. [184], and experimental data are from Refs. [145, 146]. Average speed of sound, $V_{ave} = 1/3(V_l+2V_t)$.

| Polymer | $V_l$ (m/s) | $V_t$ (m/s) | $V_{ave}$ (m/s) | $\kappa$ (W/mK) |
|---|---|---|---|---|
| PALi (MD) | 7456 | 3815 | 5029 | 0.61 |
| PANa (MD) | 6395 | 3440 | 4425 | 0.67 |
| PAMg (MD) | 6577 | 3225 | 4342 | 0.54 |
| PAAl (MD) | 6833 | 3610 | 4684 | 0.61 |
| PAK (MD) | 6624 | 3562 | 4583 | 0.51 |
| PACa (MD) | 6311 | 2936 | 4061 | 0.43 |
| PACr (MD) | 6856 | 3515 | 4629 | 0.69 |
| PAFe$^{2+}$ (MD) | 6818 | 3567 | 4651 | 0.61 |
| PAFe$^{3+}$ (MD) | 6960 | 3544 | 4683 | 0.70 |
| PANi (MD) | 6540 | 3028 | 4199 | 0.64 |
| PACu (MD) | 6407 | 3019 | 4148 | 0.60 |
| PASn (MD) | 5044 | 2525 | 3365 | 0.40 |
| PAPb (MD) | 5013 | 2470 | 3318 | 0.33 |
| PAHF (MD) | 6772 | 3512 | 4599 | 0.61 |
| PAHCl (MD) | 6678 | 4167 | 5004 | 0.32 |
| PAHBr (MD) | 5455 | 3452 | 4120 | 0.23 |
| PALi (exp) | 5100 | 3000 | 3700 | 0.55 |
| PANa (exp) | 4100 | 2500 | 3033 | 0.45 |
| PACa (exp) | 4800 | 2600 | 3333 | 0.49 |
| PVPA (exp) | 3900 | 2200 | 2766 | 0.44 |
| PVPLi (exp) | 5300 | 3100 | 3833 | 0.63 |
| PVPCa (exp) | 5300 | 3000 | 3767 | 0.67 |
| PVSNa (exp) | 4300 | 2200 | 2900 | 0.42 |
| <span style="color:red">PAA (MD)</span> | 4616 | 2258 | 3044 | 0.31 |
| <span style="color:red">PS (exp)</span> | 2380 | 1120 | 1540 | 0.14 |
| <span style="color:red">DSQ (exp)</span> | 2100 | 740 | 1193 | 0.14 |
| <span style="color:red">PC71BM (exp)</span> | 3000 | 1360 | 1907 | 0.06 |
| PMMA (exp) | -- | -- | 1809 | 0.20 |
| PVA (exp) | -- | -- | 2023 | 0.31 |
| PAA (exp) | -- | -- | 2392 | 0.37 |
| PVP (exp) | 3180 | 1330 | 1947 | 0.27 |
| PAM (exp) | 4340 | 1820 | 2660 | 0.38 |
| PSS (exp) | 3640 | 1300 | 2080 | 0.38 |
| MC (exp) | 2770 | 1150 | 1690 | 0.21 |
| PAP (exp) | 2640 | 1300 | 1747 | 0.16 |
| PAA (cross-linked) (exp) | 3450 | 1640 | 2243 | 0.28 |

### 3.5. CHEMISTRY EFFECT

As discussed previously, while polymers share similar low temperature thermal conductivity behavior since they all behave like an elastic medium to support long wavelength transport, their higher temperature thermal conductivity becomes diverse in temperature dependency and amplitude. The above discussion has eluded such differences to the inherent chemistry of different polymers. In our opinion, the ability to correlate chemistry of polymers and their thermal conductivity is the crown-jewel of this field since it would provide the ultimate guidance needed for materials by design. Chemistry of the polymer directly determines the atomic mass and bonding natures, which further influences the conformation of chains, density, the ability to form crystal structures, interatomic interaction, heat capacity, speed of sound, etc. As mentioned in Hands et al.'s work in 1973,[151] research efforts had aimed to provide chemists and engineers with assistance to predict the thermal conductivity of a polymer given their chemical nature, such as molecular weight, degree of branching and crosslinking, stereoregularity, crystallization, defects (voids and structural irregularities), and molecular orientation, together with external parameters like temperature and pressure. It is fair to say that important physics has been understood along the way, but the ability to precisely predict thermal conductivity based on chemistry still needs much work. However, we are seeing increased efforts in recent years to achieve understanding of the correlation between the chemical composition of a polymer and their bulk thermal conductivity. In this section, we discuss works that have helped unravel the chemistry-thermal conductivity relation.

### 3.5.1. CHAIN LENGTH EFFECT

Relating thermal conductivity to even the simplest chemistry feature, the chain length (i.e., degree of polymerization), has not been straightforward. Early experiments by Hansen et al.[194] on PE found that the thermal conductivity of amorphous polymers scaled with the square root of the molecular weight for short chains and converges when the chain lengths are sufficiently long. The observed relation was $k \propto \sqrt{M_w}$ when $M_w < 100,000$. Since then, this trend of thermal conductivity increasing with $M_w$ was also observed in other types of polymers, such as PMMA,[195] polystyrene (PS),[196] and polycaprolactam (PCL).[197] For sufficiently long chains, Fesciyan et al.[198] derived a model for thermal conductivity based on the Green-Kubo formula for high molecular weight polymer melts, but the derived formula was not a function of chain length. However, for shorter chains, models show diverse trends. The first model in the literature that relates thermal conductivity to molecular weight is Weber's equation for liquids,[168] $\kappa = M_c c_p M_w^{\frac{1}{3}} \rho^{\frac{4}{3}}$, where $M_c$ is related to the material property, $c_p$ is the specific heat at constant pressure, $M_w$ is the molecular weight and $\rho$ is the mass density. According to this equation, thermal conductivity would scale with the chain length to the 1/3 power since molecular weight is linear to the degree of polymerization for the same polymer. This power, however, deviates from the experimentally observed $\kappa \propto \sqrt{M_w}$ for short PEs.[194] There were also other models suggesting an inverse relationship against molecular weight, $k \propto (\frac{1}{M_w})^\alpha$, where $\alpha$ was 0.3 (Ref. [199]) or 0.5 (Ref. [200]). These examples show the difficulty in developing a physical understanding that can relate the thermal conductivity to the polymer chemistry in amorphous polymers.

MD simulations provide unique opportunities to gain insights from the molecular level for polymer thermal conductivity. Ohara *et. al.*[201] used MD to calculate the thermal conductivity of short PEs (n-alkane) with chain length up to 24 carbon segments, but there was no obvious trend as a function of chain length observed. In this study, they developed a method to decompose the thermal conductivity into contributions from different inter-atomic interactions (bonding and non-bonding interactions) and molecular advection. They found that as chain length increases, the contribution of intra-chain bonding interaction increases monotonically and becomes larger than non-bonding interactions. For $C_{24}H_{50}$, intra-chain bonding forces contribute ~ 50% of total thermal conductivity, compared to ~30% from non-bonding interactions and ~20% from advection. This is an important result since it implies that thermal transport along the chain can be as important as, if not more important than, non-bonding interactions – a critical difference between polymers, simple liquids (little intra-chain contribution) and inorganic amorphous materials (purely bonding interaction). Another study from Ohara's group[202] found that the thermal conductivity of PAA increases with chain length due to the enhanced intra-molecular interactions. The same group[203] later compared linear alcohol and linear alkane with different chain lengths, and found that thermal conductivity of alcohol was uniformly larger than alkane for the same chain lengths, but the thermal conductivity of both materials seems to converge as chain length increases. This was attributed to the contribution from the polar hydroxyl end groups in alcohol, which enhanced inter-molecular interaction. As the chain length increases, the role of end groups decreases and thus the two thermal conductivity results converge. Zhao et al.[204] studied the thermal conductivity dependence on chain length in amorphous PE using MD simulations with the chain length from 4 to 1260. The calculated thermal conductivity increases initially and then reaches a plateau at higher degree of polymerization – the same trend seen in experiments. They found that there was a clear correlation between morphology and thermal conductivity, and the major differentiator is the phase of the polymer (gas, gas-liquid, and liquid) due to different chain lengths.

Using MD simulations, Wei et al.[205] studied the chain length effect on the thermal conductivity of amorphous PE, with the degree of polymerization ranging from 5 to 200. The thermal conductivity was found to scale with $\kappa \sim L^{0.44}$, where $L$ is the chain length, which is close to the square root relation from experiments.[194] By decomposing the contribution to thermal conductivity from advection, non-bonding and bonding interactions, and describing each of the contribution using existing models or newly created ones, a thermal conductivity-chain length relation considering density, chain conformation (described by $R_g$) and chain stiffness was proposed. The model was shown to agree with the MD results well. These studies focusing on a simple chemistry feature, chain length, showed that there are many factors (e.g., density and chain conformation) that are the results of the chemistry that can impact thermal conductivity.

### 3.5.2. CHAIN CONFORMATION EFFECT

Some recent studies have pointed out that polymer chain conformation is a key factor influencing thermal conductivity even if the materials are in purely amorphous states. These are mainly understood by detailed MD simulations. As mentioned previously, an important feature of polymers compared to inorganic amorphous materials is their local anisotropic bonding environment consisting of both covalent bonding and non-bonding interactions like vdW and electrostatic interactions. It has been shown that in amorphous polymers, thermal transport along the chain backbone via the strong covalent bonds contributes more than those non-bonding interactions to thermal conductivity (see **Fig. 12b**). In recent MD simulations, Luo and coworkers have shown that the thermal conductivity of amorphous polymer is closely related to the chain conformation.[111, 128, 206] It has been observed that thermal transport along the covalent chain backbone is related to the chain conformation, especially its spatial extension as characterized by the radius of gyration ($R_g$).[111] Through a parametric study using MD simulations, the $R_g$ of a model PE is systematically tuned by changing its dihedral angle energy constant.[111] As the dihedral angle is strengthened, there will be less segmental rotation along the chain and then the persistence length of the chains becomes larger, which in turn lead to larger $R_g$ (**Fig. 15a**). While there is no direct proof, it is reasonable to think that larger $R_g$ would allow heat transfer along the chains to reach longer spatial distantances before getting interrupted by chain ends. As seen in **Fig. 15b**, the PE with larger $R_g$ has larger thermal conductivity contributed from the intra-chain bonding interactions. Such a finding indicating a positive correlation between $R_g$ and thermal conductivity was not only observed for homopolymers,[111] but was also generalized to polymer blends[128] and bi-/tri-block copolymers[206] (**Fig. 15c**).

Stiff polymer chains may help identify polymers with high thermal conductivity. In reality, π-conjugated polymers usually have stiffer backbone.[56] An experimental study by Singh *et al*.[7] shows that polythiophene (PT), a π-conjugated polymer, can have a large thermal conductivity (~4 W/mK) in the amorphous, but somewhat aligned, state. This was attributed to the stiff backbone of PT, which keeps the chain straight over a long distance. In another study by Xu et al.,[104] it is shown that the strong backbone of poly(3-hexylthiophene) leads to a high thermal conductivity of 2.2 W/mK in an oxidative-CVD synthesized amorphous film. In addition to the strong backbone, the authors also argued that the strong π-π stacking facilitated inter-chain thermal transport. In a recent simulation study (not yet published), we observe that the π-π stacking does not directly enhance thermal transport across chains, but instead, it helps straighten chains and lead to enhanced intra-chain thermal transport along the covalent backbone.

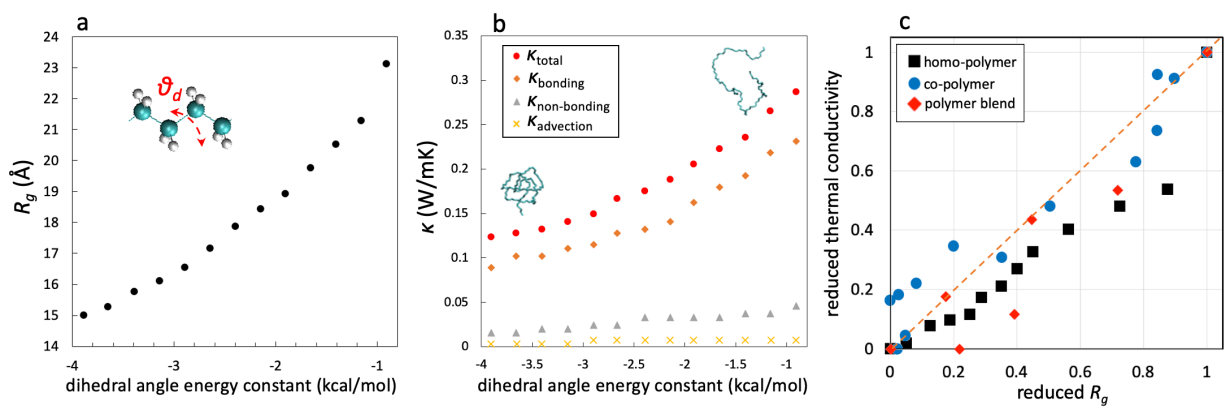

**Figure 15.** (a) Parametric study in MD simulation shows a positive relation between $R_g$ and dihedral angle energy constant. (b) Thermal conductivity decomposition as a function of the dihedral angle energy constant. (c) The relationship between reduced thermal conductivity and reduced radius of gyration of different types of polymers in amorphous states. Reduced thermal conductivity $\underline{\kappa} = \frac{(\kappa - \kappa_{min})}{(\kappa_{max} - \kappa_{min})}$, where $\kappa_{min}$ and $\kappa_{max}$ are respectively the minimal and maximum thermal conductivity in each type of polymer. Similarly, reduced radius of gyration $\underline{R_g} = \frac{(R_g - R_{g,min})}{(R_{g,max} - R_{g,min})}$.

Of course, the chain conformation effect on thermal conductivity can become more complicated when the complexity of the polymer chemistry increases, such as side chain decoration.[207, 208] For example, our prior laser pump-probe experiments[208] using a time-domain thermoreflectance (TDTR) system showed that the thermal conductivity of a series of amorphous conjugate polymers, PBDTTT, can be manipulated when the side chains are modified even though the backbones are exactly the same (**Fig. 16a**). The polymer with

Thermal Transport in Polymers: A Review. 

linear side chains (PBDTTT-DD) exhibits a thermal conductivity that is 160% higher than that with shorter and bulkier branched side chains (PBDTTT-EE) (**Fig. 16a**). The SAXS and gazing incident X-ray scattering (GIXS) characterizations show that polymers with long side chains tend to have higher crystallinity and larger crystal sizes, but the reason is not currently clear.[208] When such factors are considered, we were able to explain the thermal conductivity difference between the DD and the rest types of polymers using the effective medium approximation[143, 209] (squares, **Fig. 16b**). However, it was not until we also considered the inter-molecular distance separated by the side chains (d-spacing, **Fig. 16b**) in the amorphous phase that we were able to reproduce the thermal conductivity trend over the whole range (triangles, **Fig. 16b**). This implies that not only can global morphology influence thermal transport but, in amorphous polymers, the local molecular morphologies are also important. Much fundamental research is still needed to completely understand the chain conformation effect on thermal conductivity. This will be a great challenge in this field and will be highly rewarding since it may be the critical stepstone to achieving designing thermally conductive polymers with rational chemistry selection. Tackling such a challenge, in our opinion, will require a strong collaboration between experiments, characterization, molecular simulations and potentially machine learning techniques.

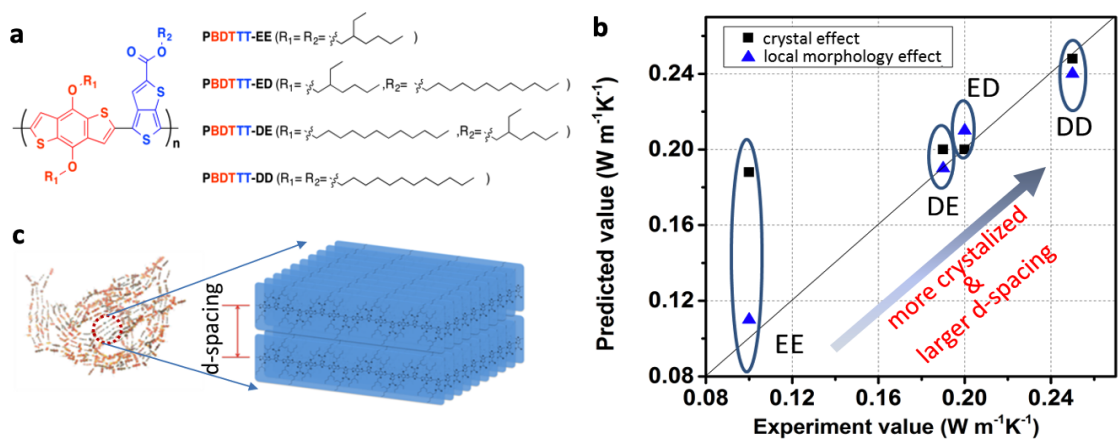

**Figure 16.** (a) Molecular structure of the PBDTTT family of polymers; (b) Comparison of thermal conductivities predicted by EMA and measured by TDTR; (c) Local structure of PBDTTT from MD

simulations.[24] D-spacing is the inter-molecular distance separated by the side chains.

### 3.5.3. POLYMER BLENDS

According to effective medium theory, in a binary mixture of two polymers the thermal conductivity would vary monotonically as the fraction of one component changes. For example, Morikawa et al.[210] showed that the thermal diffusivity of poly(phenylene oxide) (PPO) and polystyrene (PS) binary blends increase almost linearly from 1.13 to 1.63 × $10^{-7}$ $m^2$/s, when the PPO ratio increases from 0 to 100 $_{wt}$%. Another study from the same group[211] also showed that polymer blends of poly(ethylene-co-cyclohexane 1,4-dimethanol terephthalate) (PETG) and bisphenol-A polycarbonate (PC) had a monotonic trend when the PC content increases, although the trend was not linear. A recent report[212] showed that the thermal conductivity of a blend made of PC and ethylene-propylene copolymer (EPC) with 10 $_{wt}$% PC was ~ 0.206 W/mK, between that of PC (0.240 W/mK) and PE (0.166 W/mK). Duda et al.[213] studied the thermal conductivity of P3HT:PCBM blends, and found it to have a linear relationship with the P3HT concentration.

There are also indications that the thermal conductivity in polymer blends cannot be simply described by the effective medium theory. In 2014, Guo et al.[214] observed that the binary mixture of a fullerene derivative and a conjugated polymer had a minimum thermal conductivity of ~0.06 W/mK at 30-35 $_{vol}$% of the conjugated polymer, which was even lower than that of the pure fullerene derivative (~0.07 W/mK). This observation was potentially related to phase segregation, which has not yet been fully understood. In 2016, Kim et al.[10] reported that the polymer blend of poly(N-acryloyl piperidine) (PAP) and poly(acrylic acid) (PAA) could reach a thermal conductivity as high as 1.5 W/mK at 30% molar fraction of PAP. This high thermal conductivity was unprecedented and was about an order of magnitude higher than that of common amorphous polymers. The mechanism was explained as that the interchain hydrogen bonds between PAP and PAA molecules created a homogeneous network that enhanced thermal transport. It is worth mentioning that the authors also observed non-monotonic relation between thermal conductivity and blending ratios for another two blends, PAP-PVA and PAP-PVPh. The thermal conductivity of these two blends (< 0.4 W/mK) were much smaller than that of the unique PAP-PAA blend. On the other hand, an independent research by Xie et al.[146] reported that PAP and PAA could not form a homogeneous blend, and that PVA-PAA blends did not show an extraordinary thermal conductivity value at intermediate mixing ratios, even though they contained some interchain hydrogen bonds. Even if a homogeneous hydrogen bond network could form, the extremely high thermal conductivity of ~1.5 W/mK would be surprising, since in a similar vein, cross-linking, which forms homogeneous covalent bond networks led to even lower thermal conductivity for PAA (from 0.37 to 0.28 W/mK).[146]

Recently, Bruns et al.[215] used MD simulations to find that PAP and PAA were immiscible at any mixing ratio, and that the PAP-PAA blend showed almost invariant thermal conductivity as the mixing ratio changed. They also reported that polyacrylamide (PAM)-PAA blends had a maximum thermal conductivity of ~0.42 W/mK at the 40% monomer molar fraction of PAM, which was attributed to that the PAM created more bridges for the major polymer PAA through interchain hydrogen bonds. However, this argument can also be countered by the lack of thermal conductivity improvement in cross-linked PAA.[146] Bruns et al. also believed that the thermal conductivity improvement by blending PAM and PAA was related to the increase in stiffness (bulk modulus) due to the hydrogen bond network. However, another experimental study[216] showed that due to the interchain hydrogen bond network formation by adding water, the thermal conductivity of PVA and its blends could be enhanced from 0.3-0.4 W/mK to ~0.7 W/mK, despite the significant decrease in modulus after these polymers were moisturized. Wei et al.[128] studied the effects of interchain and intrachain interactions between the blending polymers by artificially tuning the interchain and intrachain interaction strengths in a model MD simulation. It was found that increasing the interchain interaction could enhance thermal conductivity, but it was due to the polymer chains in the major phase being stretched (i.e., $R_g$ increased) and the heat flux through bonding interaction being enhanced.

For polymer blends, different and sometimes contradicting results have been reported, and the explanation of thermal transport physics is under significant debate. There are some immediate questions worth answering. For example, can blending reform the polymer chain conformation, and if such an effect has an obvious impact on thermal conductivity? Much work combining experiments, characterization and MD simulations is needed to advance our understanding for thermal transport in polymer blends.

### 3.5.4. POLYELECTROLYTES

The above text focused on charge neutral polymers. The charges and highly polarized groups in polyelectrolytes, which result in strong Coulombic interactions, add another level of complexity to the thermal transport mechanism. In polyelectrolytes, due to the strong Coulombic interactions, the electrostatic forces can directly enhance thermal transport via stronger inter-molecular interactions or it may enhance thermal transport by changing the chain conformation. Shanker et. al.[217] showed that at high pH the ionized PAA can reach a thermal conductivity value up to ~1.2 W/mK, and they suggested that such an enhancement is related to the electrostatic interaction between the polarized groups on the same polymer chain backbone helping stretch the polymer backbones, which in turn increase the thermal transport along the chain – similar to the effect of enhanced $R_g$. However, Xie et. al.[145] found that the highest thermal conductivity of various polyelectrolytes, including PAA, was not more than ~0.67 W/mK. According to the counterion condensation theory, bulk amorphous polyelectrolytes tend to have a collapsed chain conformation, i.e., $R_g$ would decrease.[218-220] This thus contradicts with the mechanism proposed by Shanker et. al.[217] In recent MD simulations,[169, 184] it was found that thermal conductivity would indeed increase as the ionization increases (from ~0.30 to ~0.67 W/mK, **Fig. 17a**), with the level of increase close to that found in experiments.[145, 217] However, the simulations failed to find obvious increase in $R_g$ as a function of ionization, but instead found that the thermal conductivity increase could be largely attributed to the counterion-polymer electrostatic interaction. Even if with an artificial increase in the $R_g$ of fully ionized PAA from ~12.3 to ~16.1 Å through increasing the dihedral angle strength, the thermal conductivity only slightly increased from ~0.67 to ~0.70 W/mK.[169]

While as the ionization level increases lead to obvious enhancement in the Coulombic interactions (**Fig. 17b**), it was interesting to find that the thermal conductivity increase was instead mainly due to the contribution from Lennard-Jones (LJ) interactions (**Fig. 17c**). The Coulombic force does not directly contribute to heat transfer, but it attracts oppositely charged atoms closer, which in turn increases LJ forces between significantly, especially the LJ repulsive force ($\propto r^{-12}$) (**Fig. 17d**). The molecular-level understanding is that the enhanced Coulombic interaction between the ionized polymer functional groups and counterions attracts them closer together, and the LJ interaction transitions from attractive to repulsive (**Fig. 17e**), which is evident from **Fig. 17b** showing the LJ potential changes from positive to negative when ionization level increases. Since the repulsive portion of the LJ potential is much steeper than the attractive portion (i.e., larger force amplitude) (**Fig. 17e**), being in the repulsive region would greatly enhance thermal conductivity which can be expressed as the product of the interatomic force and the atomic velocity in the molecular level.

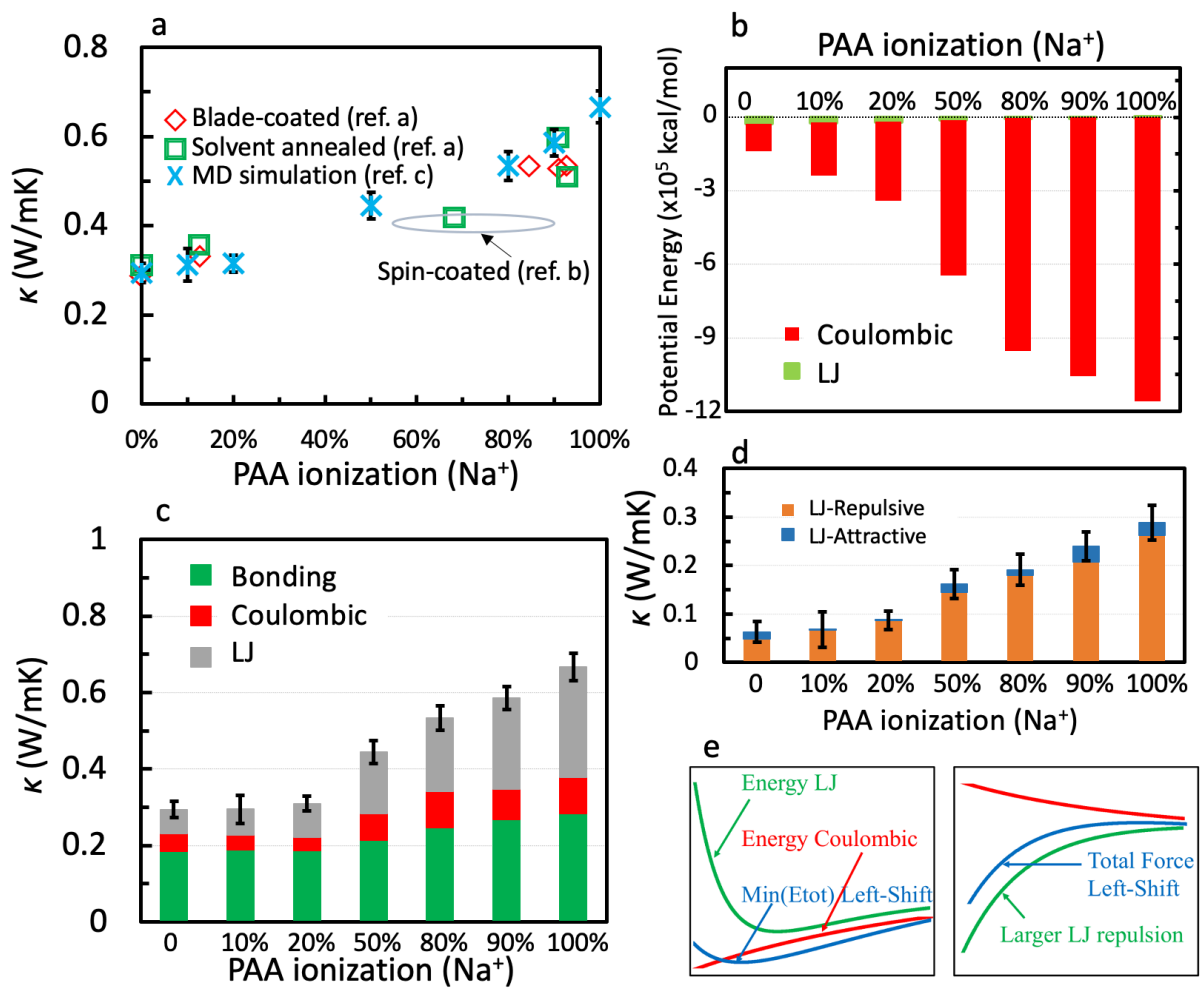

**Figure 17.** (a) Thermal conductivity of PAA as a function of ionization level with Na$^+$ as the counterions. Both experimental results (ref. a:[217], ref. b:[145]) and MD simulation results (ref. [169]) agree well with one another. (b) Non-bonding interactions, including Coulombic and LJ interactions, as a function of ionization level. (c) Thermal conductivity decomposition into contributions from bonding interactions and non-bonding interactions including Coulombic and LJ. **(d)** Thermal conductivity contributed from attractive and repulsive LJ interactions. (e) Schematics showing (left) how Coulombic interaction attract atoms closer together and shifts the interatomic distance to the repulsive region of the LJ potential, and (right) how the repulsive force dominants interaction in the repulsive region.[169]

In polyelectrolytes with different types of counterions, the ionic charges and radii are different, which directly impact the counterion-counterion, counterion-polymer and polymer-polymer interactions. These will further influence thermal conductivity. Both experimental study[145] and MD simulation[184] show that thermal conductivity of common polyelectrolytes are all < 0.7 W/mK. Based on the MD simulation results, using a machine learning tool, Random Forest, the feature importance of different descriptors (atomic mass, atomic radii, vdW radii and ionic radii) associated with the counterions was quantified.[184] The ionic radius emerged to have the strongest relationship with thermal conductivity, although PAA itself was an outlier (**Fig. 18a**). This finding might be interpreted as that when the ionic radius increases, the pairwise force decreases, and as a result, the thermal conductivity decreases. With physical reasoning, a combined parameter, $n^2 F_{LJ} d_{RDF}/\sqrt{m}$, which was derived from the molecular-level heat flux, was proposed. This combined parameter involves the equilibrium LJ force ($F_{LJ}$), atomic mass ($m$), number density ($n$) and RDF peak location ($d_{RDF}$), and can describe a positive relationship between counterions and thermal conductivity (**Fig. 18b**). Such a combined parameter shed the light on the complexity of how ions could impact thermal conductivity. Thermal transport mechanism in polyelectrolytes is still not fully understood, and the possibility of making polyelectrolytes thermal conductivity > 1 W/(m.K) remains to be verified. The important aspects that can affect the thermal conductivity of polyelectrolytes are polymer chain conformation, ionization ratio (pH), ionization position, polymer backbone type, counterion type, and in practice, water content. Using molecular simulations and potentially combined with machine learning tools, there is hope to unravel the mechanism of such a complicated problem.

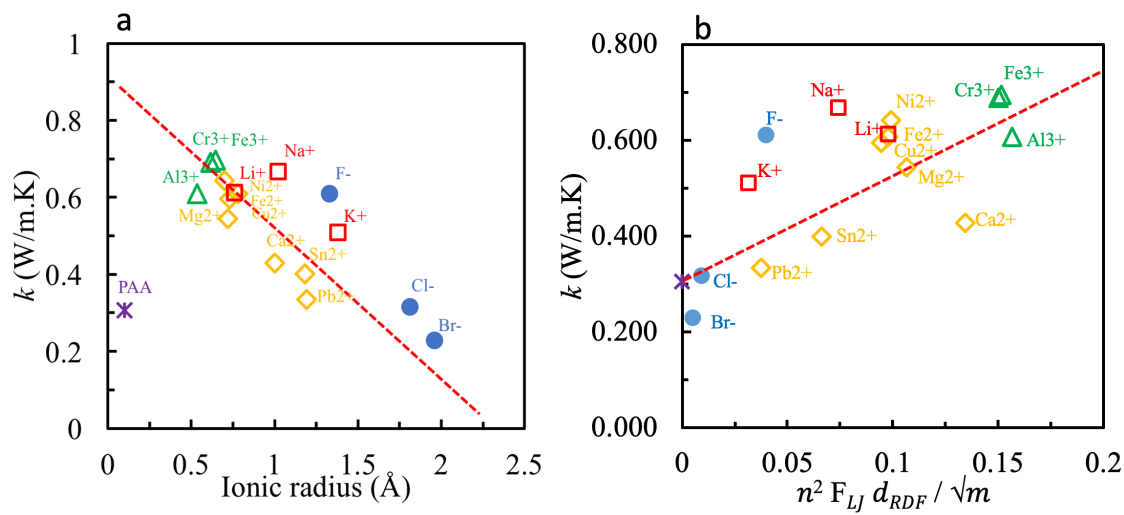

**Figure 18**. **(a)** Thermal conductivity of PAA with different counterions plotted against their ionic radii, and **(b)** a combined descriptor derived from physical reasoning.

Before we conclude Section 3.5, we would like to note that the above topics related to polymer chemistry are far from complete, and more importantly, they are not isolated but highly intertwined. For example, blending polymers can result in conformation change of the polymer chains and it can also modify the inter-molecular interactions, both of which can lead to change in thermal conductivity. We believe that the community has only scratched the surface of chemistry-thermal conductivity relation, and much work is needed to advance and deepen our understanding in this field.

## 4. THERMAL REGULATION USING POLYMERS

The ability to reversibly modulate polymer thermal conductivity can have potential applications in data storage, sensing, thermal management and thermal logics.[221, 222] Different mechanisms to modulate polymer thermal conductivity have been explored. These include controlling temperature, modulating chemical bonds and UV-light excitation. Below, we highlight some examples for each of these methods.

4.1. TEMPERATURE STIMULATION

Using MD simulations, Zhang et al.[66] has shown that thermal conductivity of a PE crystal fiber can be regulated reversibly through temperature, mechanical strain or their combinations. In crystalline PE fibers, thermal transport is dominated by propagating phonons, and their MFP and thus thermal conductivity is very sensitive to the segmental disorder along the chains. The energy constant of the dihedral angle, which controls the segmental rotation, is on the order of ~0.1 kcal/mol – ~0.8$k_BT$ for room temperature. This means that thermal energy in the intermediate temperature range would be sufficient to overcome the dihedral angle energy barrier to enable segmental rotation (**Fig. 19a**). The ground state of the dihedral angle of PE is at 180º (*trans* conformation), but there are metastable states at angles of 60º and 300º (*gauche* conformation), and the emergence of the *gauche* conformation can happen within an extremely narrow temperature window of 2 K (**Fig. 19b**). By controlling the temperature only, the population of such disorder was able to be controlled and the thermal conductivity could be switched back and forth reversibly with a ratio of ~7 between 300 and 450 K. With the help of strains, which help eliminating the *gauche* population, the switching ratio could be as large as ~10 between 300 and 450 K and ~6 between 380 and 400 K. Recent experiments by Shrestha et al.[12] based on ultra-drawn PE fibers[11] successfully realized the MD simulation results, showing record-high solid-state thermal switch with a switching ratio of 10. Such a switchable thermal conductivity of PE was further developed in MD simulations to realize thermal diode with very high rectification factors,[65] and experimental validation was recently acheived.[223]

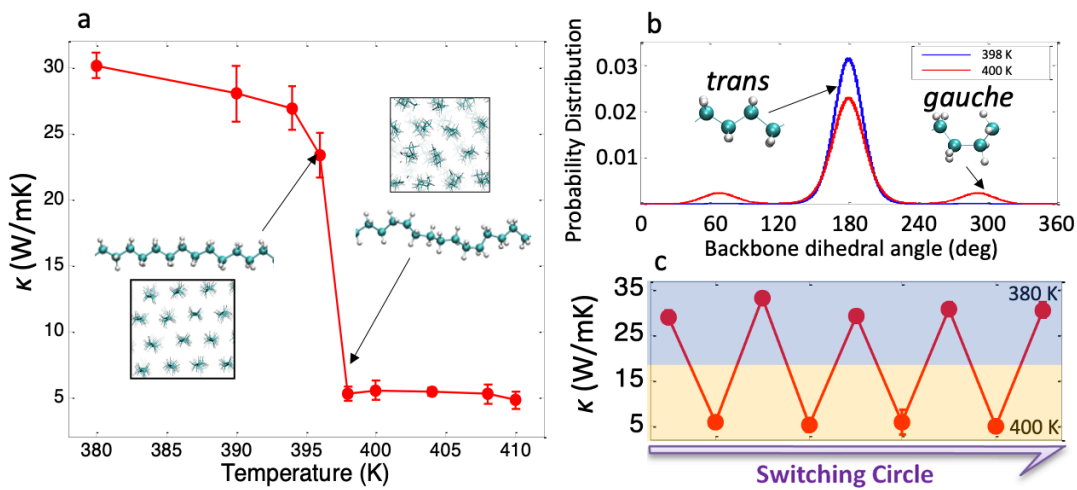

**Figure 19**. **(a)** Sharp thermal conductivity changes from 380 to 410 K due to the thermal excitation of segmental rotation. **(b)** The emergence of *gauche* conformation can happen within a 2 K window. **(c)** Reversible thermal switch between 380 and 400 K with 20% strain applied.

Temperature can also induce phase transition of bulk polymers and their composites,[224] which can also in turn lead to thermal conductivity regulation. Li et al.[225] leveraged the second-order phase transition in poly(N-isopropylacrylamide) (PNIPAM) and demonstrated that the thermal conductivity could be switched with a ratio up to 1.15. The mechanism was attributed to the contraction and expansion of the polymer chains in the aqueous solution, which could be regulated by temperature. Using the phase-change induced thermal conductivity regulation, thermal diodes have also been demonstrated.[226]

### 4.2. CHEMICAL REGULATION

Chemical method that modifies the bonding inside macromolecules has also been leveraged to regulate thermal conductivity. Tomko *et. al.*[227] demonstrated experimentally that thermal conductivity of topologically networked bio-polymers could be enhanced by ~4x when hydrated compared to the dry state (1.3 W/mK compared to 0.33 W/mK). The change in thermal conductivity was attributed to the fact that the displacement amplitude of atomic vibrations could be enhanced when hydration breaks the hydrogen

Thermal Transport in Polymers: A Review.    60

bonds between the DNA strands, allowing the strands to move more freely. The authors used the picture of Allen and Feldman[186] and argued that the transport of diffusons is directly related to the mean square displacement of atoms. When the strands move with a higher mean square displacement in the dissociated phase of the hydrated state, the diffusivity of the diffusons increases and thus thermal conductivity increases. Using transient grating spectroscopy, Li et al. observed abrupt thermal conductivity change across the lower-critical solution temperature of thermoresponsive polymer PNIPAM aqueous solution.[225] In a different study, Feng et al.[228] also leveraged the hydrophobicity change across the low critical solution temperature of the PNIPAM hydrogel to tune the internal morphology and water content for thermal regulation. They demonstrated a thermal switching factor up to 3.6. Using a polyacrylamide (PAAm) hydrogel, Tang et al. showed that the thermal conductivity can vary from 0.33 to 0.51 W/mK due to hydration-induced cross-linking and water content change, while the thermal conductivity was not sensitive to temperature in the range of 25-40 °C.[229]

### 4.3. PHYSICAL EXCITATION

Physical stimulation is very attractive for thermal regulation, since it can modulate the thermal conductivity through an external physical field. Shin et al.[230] showed that applying a magnetic field could help align liquid crystal monomers in the nematic phase, which in turn led to a thermal conductivity modulation factor of ~2. This change was related to the alignment of the liquid crystal molecules with respect to the heat transfer direction. When the magnetic field was parallel to the substrate, the thermal conductivity was 0.14 W/mK, and it increased to 0.24 W/mK when the field was rotated by 90°. Recently, another study from the same group[231] demonstrated thermal regulation using light excitation. It was shown that a light-responsive polymer could switch its thermal conductivity from 0.35 W/mK under normal condition to 0.10 W/mK upon UV-light excitation, reaching a modulation factor as high as 3.5. The mechanism was attributed to that the UV-light could change the azobenzene side-chain from *trans* to *cis* conformation. The *cis* conformation would break the $\pi - \pi$ stacking structure in the polymer, and as a result the local structure transitioned from an ordered crystal structure to a more disordered liquid structure, leading to changes in thermal conductivity. Our MD simulation results (not published) indicated that the $\pi - \pi$ stacking structure could stretch the backbone of the side-chain, and the thermal conductivity enhancement could be mainly attributed to the side-chain bonding interaction's contribution to thermal transport – an effect similar to the chain conformation effect (e.g., $R_g$).[111, 128, 206]

### 4.4. INTRINSIC THERMAL RECTIFICATION OF POLYMER MOLECULES

Besides external excitations, some special molecules process intrinsic thermal regulation properties. For example, tapered bottlebrush polymers have asymmetric polymer architecture. They consist of a linear polymer backbone with side chains of systematically varied molecular weights that can be tailored to produce a cone-shaped macromolecule. Using NEMD simulations, Ma and Tian demonstrated that these polymers have the unique ability to generate significant thermal rectification that cannot be rationalized based on conventional wisdom.[232] In sharp contrast to all other reported asymmetric nanostructures, they observed that the heat current from the wide end to the narrow end (the forward direction) in tapered bottlebrush polymers is smaller than that in the opposite direction (the backward direction). Further analysis showed that a more disordered to less disordered structural transition within tapered bottlebrush polymers is essential for generating nonlinearity in heat conduction for thermal rectification. Moreover, the thermal rectification ratio increased with device length, reaching as high as ~70% with a device length of 28.5 nm as shown in **Fig. 20**. This large thermal rectification with strong length dependence uncovered an unprecedented phenomenon−diffusive thermal transport in the forward direction and ballistic thermal transport in the backward direction. This is the first observation of a switching between different heat transfer regions as the heat flow direction flips. The fundamentally new knowledge gained from this study may spark interest into intrinsic thermal rectification in asymmetric polymer molecules. Future work to scale up the single molecule rectification would be needed to drive practical applications.

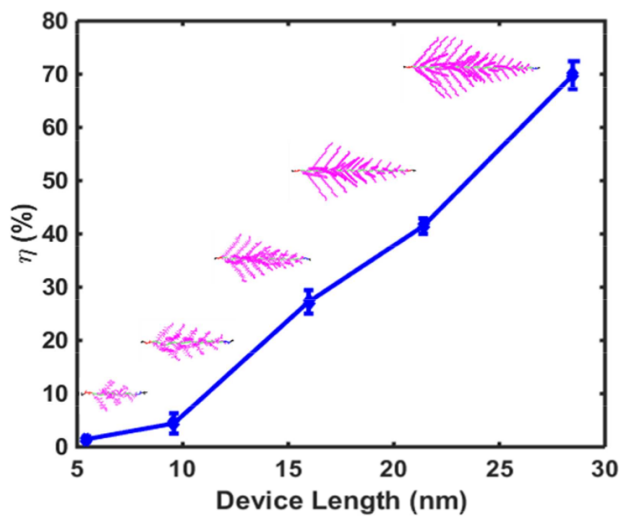

**Figure 20.** Thermal rectification ratio of tapered bottlebrush polymer versus its length.[232] Reproduced with permission.[232] Copyright 2018, American Chemical Society.

## 5. THERMAL TRANSPORT ACROSS INTERFACES INVOLVING POLYMERS

While we intentionally choose not to focus much on polymer composites to emphasize the discussion of the physics of polymer thermal transport, we would like to devote this section to briefly discuss the interfacial thermal transport related to polymers, especially between polymers and solids, which can be important to composite thermal conductivity especially when the matrix polymer thermal conductivity becomes higher (**Fig. 10**). We refer the readers to the review articles by Chen et al.,[33] Huang et al.[36] and Yang et al.[37] on the topic of polymer composite thermal conductivity.

In general, the role of interfaces becomes more important as the characteristic lengths of the materials shrink, which makes the interface density high and interfacial thermal resistance of similar magnitude as that of the constituents.[15, 30, 233, 234] Even in pure soft materials, interfaces exist between crystal regions and amorphous regions.[181]

### 5.1. INTERFACIAL BONDING EFFECT

Research on interfacial thermal transport across hard-soft material interfaces has generally focused on binding energy – a critical factors that influence interfacial thermal transport.[22, 27, 235-238] Properly functionalizing solid surfaces can enhance thermal conductance between metal and organics by up to three times due to the improved hydrophilicity.[26, 235] Forming covalent bonds at the interface can enable even larger enhancement in thermal conductance.[9, 23, 24, 239, 240] Hydrogen bonds, which are far less energetic than covalent bonds, however, were found to enhance thermal conductance across hard-soft interface by similar amplitude as covalent bonds.[28, 241] In such interfaces, the interactions consist of vdW interactions and strong Coulombic interactions between highly polar groups (e.g., hydrogen bond donors and acceptors). It would be intuitive to believe that it is the strong Coulombic interaction that led to the enhanced thermal conductance compared to non-polar interfaces. However, detailed thermal conductance decomposition into contributions from different interactions and structure characterization near the interface suggest that the enhanced conductance was a collaborative effect from vdW and Coulombic interactions. It was found that Coulombic forces attract the organic molecules closer to the interface, which in turn increased the force from vdW interactions. Interestingly, it turned out that vdW was the direct contributor to the thermal conductance enhancement while Coulombic was helping in an indirect manner.[28, 242] Such a molecular thermal transport picture is similar to that found in bulk polyelectrolytes as discussed in Section 3.5.4.

**5.2. HEAT CARRIER COUPLING EFFECT**

In crystal interfaces, vibration spectra matching is deemed to be the most important factor in interfacial thermal transport as reflected in theories like the diffuse mismatch model.[237] For interfaces between soft and hard materials, large vibrational spectral mismatch usually exist due to their distinct composition and bonding natures. The thermal conductance of such interfaces is thus usually at the low end (~O(10) W/mK).[27, 235, 243, 244] Recently, experimental measurements demonstrated a significant enhancement, by as much as 7x, in thermal transport across soft-hard interfaces (Au-PE) by coating the hard surfaces with self-assembled monolayers (SAM) (**Fig. 21a**).[31] Similar observations were made in another study.[245] The success was based on the fact that the SAM molecules chosen (alkanethiols) had vibrational spectra similar to the soft material (PE) and thus bridged the vibrational spectra mismatch between Au and PE (**Fig. 21b**).[31] Counter-intuitively, such large increases were realized despite significant decreases in the interfacial binding energy when the hard surface was functionalized.

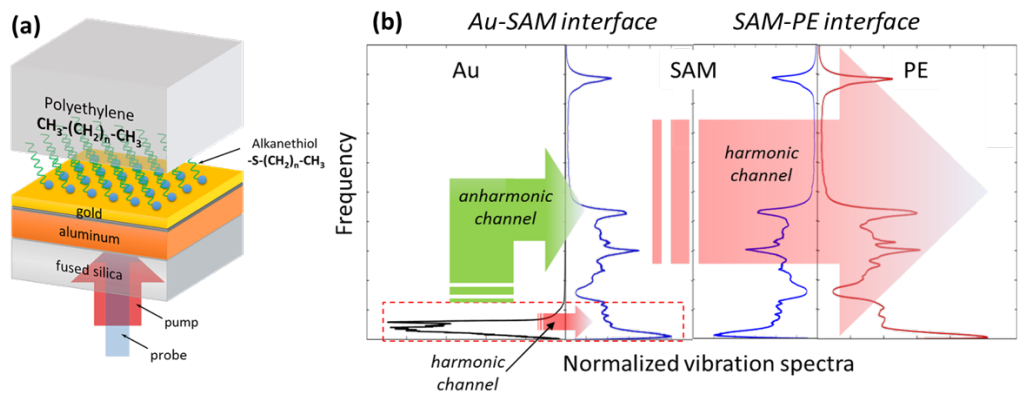

**Figure 21. (a)** Schematic of an Au-PE interface with SAM functionalization; and **(b)** vibrational spectra and hypothetical energy transport pathways across the Au-SAM-PE interfaces.

There were fundamental questions arising from such an observation. While the vibrational spectra of SAM and PE are almost identical (**Fig. 21b**), why does the Au-SAM interface have a thermal conductance much larger than that of the Au-PE interface? Conventional mismatch models[237] would have predicted similar thermal conductance for these two interfaces. One possible picture is that anharmonic channels via inelastic scattering helped distribute phonon energy from the Au to the SAM vibrational spectrum, and such energy transfer efficiency is related to the strong covalent bonds between Au and SAM. The energy transferred to the SAM spectrum from Au could then be transferred to the PE side via strong harmonic coupling due to almost perfect spectra overlap. This picture remains to be verified, and we expect non-equilibrium MD simulations coupled with spectral analysis[246, 247] to be a viable tool to tackle this problem. It is noted that in the SAM layer, thermal transport is very efficient due to the highly delocalized phonon transport along the molecular chains, which then did not add much resistance to the overall Au-SAM-PE interface.[135, 248]

Our interesting MD simulation results on a Si-PDMS interface showed that when the surface of a silicon substrate is converted from a crystal to an amorphous structure (**Fig. 22a**, the interfacial thermal conductance could be enhanced (**Fig. 22b**). Further analyses indicated that the change in surface morphology did not really change the vibrational spectra of the local atoms much (inset in **Fig. 22b**). It is possible that the nature of the heat carriers will change from propagating phonons to modes like diffusons when the Si surface is changed into amorphous structures, and such a heat carrier nature is more similar to that in the amorphous PDMS, which somehow lead to better energy exchange across the interfaces. However, this hypothesis has not yet been tested and further study is needed. We note another recent MD simulation also showed that nanoparticle morphology could influence its thermal transport to surrounding liquid, and they attributed the impact to the coordination number of surface atoms of the nanoparticle.[249]

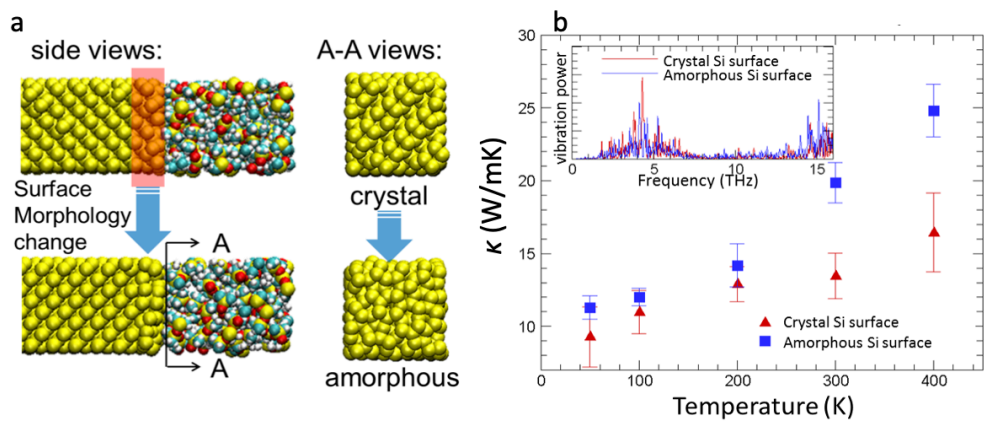

**Figure 22. (a)** Crystal (top) and amorphous (bottom) morphologies of the Si surface in a Si-PDMS interface. **(b)** Thermal conductance of these two interfaces at different temperatures.

Vibrational spectra overlap only shed light on elastic (harmonic) energy transport across the interface. However, anharmonicity, which has been largely ignored, is a potentially important factor in interfacial thermal transport. It is well-known that the increase in interfacial thermal conductance at high temperatures, where heat capacity is saturated, is because of more anharmonic channels excited at elevated temperatures. The anharmonic effects across hard-soft interfaces have also been implied from many MD simulations.[246, 250, 251] The results shown in **Fig. 22b** also implies that anharmonicity is playing a role in the observed enhancement, which becomes larger at higher temperatures. However, these studies are largely phenomenological. Models[252, 253] have been proposed to include anharmonic channels (inelastic scattering) for predicting interfacial thermal conductance, but they are significantly simplified. For example, the only inelastic channels considered were the processes that $n$ phonons with the same frequency $\omega$ scatter simultaneously to generate a phonon with frequency $n\omega$. In recent years, MD simulations on solid-solid interfaces have pointed to the fact that scattering of heat carriers prior to reaching the interface can help redistribute phonon energy to a state that is preferred for harmonic coupling across the interface.[254-256] Such pre-interface scattering is not limited to anharmonic scattering[254] but also defect scattering.[255] Could the SAM layer and the amorphous surface in the above-discussed examples have played similar roles in enhancing thermal transport across hard-soft interfaces? Further studies are needed to answer this question. MD simulation remains to be the ideal tool to answer such a question since it inherently includes anharmonicity, structure disorder and defects. Recent development on atomic Green's function to include anharmonicity[257] may also contribute to resolving the puzzle if the force constants of the hard-soft interface can be properly determined.

Other factors, such as molecular penetration,[258, 259] solid crystal orientation,[260] and local molecular re-structuring,[261] in thermal transport across hard-soft interfaces has also been studied, but we would like to refer the readers to a more detailed review of hard-soft interfacial thermal transport in the literature.[30]

## 6. PERSPECTIVE

The above text has discussed historic and recent achievements related to the understanding of thermal transport in polymers in different morphologies. It has also alluded to a number of unanswered fundamental questions and future research needs. In this section, we offer our perspectives on several directions we believe worth further research to advance this field.

**6.1. NATURE AND TRANSPORT OF HEAT CARRIERS IN POLYMERS**

A detailed study of heat carrier modes and their transport properties are still to be done to better understand the thermal transport physics in polymers. While Allen and Feldman[186, 262] have proposed a physical picture of heat carriers in disordered materials, and recent simulation and experimental studies have furthered the understanding,[189, 263-267] their theory has rarely been applied to polymers. The only application of the Allen-Feldman picture to polymers we are aware of is from Shenogin et al.[192] on amorphous PS. It showed that hardly any modes were propagons and modes above 5 THz were all locons, although the whole vibrational spectra could extend to ~90 THz. The large population of localized modes were attributed to the rigidity of PS chains, which would result in relatively high frequency vibrations. The thermal conductivity obtained from the Allen-Feldman theory were significantly lower than MD simulation results at temperatures above 100 K, and even the trends did not agree with one another.[192] Anharmonicity was considered as the possible factor for such a discrepancy, but the authors also noted that it was puzzling to find the MD yielded lower thermal conductivity than the Allen-Feldman theory, which is purely harmonic. Indeed, Alexander et al.[187] described the transport of non-propogating modes via an anharmonic coupling picture, and Lv et al.[174] used MD simulation to show that locons could contribute to thermal transport through anharmonic coupling with diffusons. While these were for inorganic materials, a similar physical picture could apply to polymers, which of course, needs verification. The ideal tool would be normal mode analysis[188] in MD simulations which inherently include anharmonic effect and captures realistic polymer structures and interactions. However, challenges exist in such studies for polymer, since in modal analysis, atomic vibration needs to be projected on to the eigenvectors of the normal modes that are determined from diagonalizing the dynamical matrix constructed from interatomic force constants at the ground state of the system. The ability of using modal decomposition for inorganic material lies in the fact that all atoms are vibrating around their equilibrium throughout the course of simulation, and thus the eigenvectors, determined from their ground state coordinates, do not change over time. For polymers, however, this could become very tricky since it is very difficult to relax the atoms to their global ground state since the movements of atoms are highly restricted by the bonding and non-bonding interactions as well as stereo-hinderance. As a result, when doing lattice dynamics, it is very likely that atoms are in metastable positions and some eigen-frequencies are negative. This would impose questions on the validity of the eigenvector to which the atomic coordinates need to be projected onto. In addition, during the simulation where atomic trajectory is tracked, the atoms can drift significantly, moving away from the initial equilibrium position. Normal mode analysis is inherently based on perturbation theory, and such large movement away from the equilibrium may invalidate this assumption meant to deal with small perturbations. Actually, according to the two-level system picture from Anderson et al.[162] and Phillips[163], it is indeed very likely that atoms would hop between metastable states back and forth (see inset schematic in **Fig. 11a**).

In addition, a key difference between polymer and inorganic amorphous materials is the locally anisotropic bonding. How would the modes transfer energy within the polymer chain through the strong bonding and across chains via weaker non-bonding interactions would be a unique question for polymer thermal transport. It would be intuitive to think that the diffusons transfer energy more efficiently along the covalent backbone and less so across chains, but why cross-linking, which forms covalent bonds between chains, turned out to reduce thermal conductivity (e.g., PAA – 0.37 W/mK to cross-linked PAA – 0.28 W/mK[146])? Is it possible that the passing of the diffusons is accompanied by the anharmonic energy communication with locons along the way? If propagons are not important in thermal transport in polymers as found by Shenogin et al.[192], would the use of $\kappa = cvl/3$,[153] or its different variants be questionable? How much can we truly borrow from the understanding of thermal transport in disordered inorganics[267] to explore polymer thermal transport? We believe these are only a fraction of the fundamental questions that are worth answering, but tools do not yet exist to fully explore them.

## 6.2. DATA-DRIVEN EXPLORATION

Besides the motivation to pursue the fundamental science, an overarching goal of studying thermal transport physics in polymer is to provide engineers with a reliable tool to predict the thermal conductivity of a polymer given its chemical nature, or to give chemists and material scientists the guidance to design a polymer based on the needed thermal conductivity to realize the so-called material-by-design. We are currently far from attaining these capabilities, although some pieces of advice can be extracted from past studies. There are so many factors, including polymer chemistry, morphology, processing conditions and etc., that can impact thermal conductivity. Obviously, establishing a model that links all these parameters to thermal conductivity is impractical given that the physics is not fully elucidated.

Materials Informatics using machine learning (ML) techniques is an important component towards the eventual realization of material-by-design.[268] ML has been recently applied to thermal conductivity,[269] especially for designing inorganic semiconductors with low thermal conductivity, mainly motivated by the demand of high performance thermoelectrics.[270-273] However, Materials Informatics concerning polymers (i.e., Polymer Informatics) has been hindered by the lack of data in a unified format.[274] For example, the largest open polymer database, PolyInfo, only has < 80 thermal conductivity data, let alone that these data contain uncertainties. Unlike inorganic crystals which have well-defined lattice structures, it is also challenging to produce large datasets for amorphous polymers using high-throughput simulations due to the uncertainty and difficulty in generating reasonable amorphous structures. Polymer has not yet been a focus of national and international initiatives like the Materials Genome program, making the data growth slow in the foreseeable future.

In a recent study, ML techniques was used to construct a surrogate model between polymer chemistry represented by SMILES (Simplified molecular-input line-entry system) strings and thermal conductivity.[275] Interestingly, they were able to obtain a deep neural network (DNN) model with reasonable accuracy ($R^2$=0.73) using merely 28 training data points in thermal conductivity available from the PolyInfo database. This was realized by implementing a novel technique called Transfer Learning, which "transfers" knowledge learned from training against highly populated proxy labels (in their case, ~6000 data points on glass transition temperature, $T_g$) into constructing DNN for sparsely populated labels (*i.e.*, thermal conductivity). By directly training a DNN against the 28 thermal conductivity data, $R^2$ was -0.4 (*i.e.*, no prediction power), but Transfer Learning helped improve the model accuracy to $R^2=0.73$. The essence of Transfer Learning is that different properties are different expressions of the inherent chemistry of materials. However, the transferred DNN is still much inferior compared to the model for the highly populated label, $T_g$ ($R^2=0.92$). It is our belief that Transfer Learning can further improve the model accuracy if more thermal conductivity labels can be leveraged in the training, and this has been shown in a recent work on inorganic crystals.[276] Some authors of the present review article are currently dedicating efforts to generate a large thermal conductivity database through high-throughput MD simulations.

In addition, $T_g$ might not be the only proxy label that can be used for Transfer Learning. Other intermediate labels, such as $R_g$, modulus, effective charges, and even force field parameters,[277] which can be calculated more quickly than thermal conductivity, may also provide a bridge to construct more accurate thermal conductivity surrogate models *via* Transfer Learning. Besides ML models, recent findings showed that properly representing polymer molecules can be important for constructing accurate surrogate models, since representations describe the chemistry of molecules.[278] It was found that a deep learning representation scheme (*i.e.*, Mol2Vec) based on the Natural Language Process (NLP) algorithm was better than the conventionally used Morgan Fingerprint, a one-hot encoding scheme based on the simple chemical connections. Even if a chemistry-thermal conductivity relation can be identified from ML, properly including processing conditions as descriptors could be another obstacle for eventually validating the model prediction, as processing conditions can impact the morphology which in turn influence thermal conductivity. Processing conditions are not something that can be easily modeled, either.

These are just a small portion of the important aspects of polymer informatics. It is our hope that there will be a collective effort from different communities (e.g., thermal, polymer, ML) to take on this problem in a holistic manner to simultaneously explore different aspects of this field (*e.g.*, database, molecular representation, ML models, MD modeling, and experimental validation) to advance it. If a chemistry-processing-property relationship is eventually established, the benefit would be beyond realizing material-by-design, as such a relation can also facilitate identifying the most influential parameters (e.g., hydrogen bonds, $\pi$-$\pi$ stacking, conjugation) that impact thermal conductivity. This would guide more targeted experimental and simulation studies to reveal specific aspects of the underlying physics.

### 6.3. MULTI-FUNCTIONALITY

The intent to use polymers for heat transfer application is not because they can conduct heat better than metals or semiconductors, but instead, is motivated by their other superior properties compared to these better thermal conductors. For example, plastic heat exchangers are more chemically resistive, lighter-weight and cheaper than metallic ones. Polymers are softer than metals so they are used as thermal interface materials to filled the gaps between rough surfaces in electronic packages.[2] Polymers are ideal for solid-state electrolytes, improving their thermal conductivity helps heat dissipation and thus safety.[279] Because of the application requirements, designing polymers with desired thermal conductivity is always constrained by their other properties. For example, loading metallic fillers into polymer matrices for thermal conductivity improvement limits their application in electronics due to short circuiting concerns. Adding excessive fillers also hardens polymers, impairing their ability to fill gaps. Materials informatics with multi-objective optimization can be a valuable tool to balance different design targets for different applications.[280]

Besides tuning polymer thermal conductivity as a secondary property for given applications, we have also recently seen studies taking advantages of the unique thermal conductivity of polymers to realize new functionalities. Thermal regulation using polymers is certainly one such example that leverages the high sensitivity of thermal transport to polymer morphologies (see Section 4). Thermally conductive polymer fibers are also being explored for wearables to help body temperature regulation, energy harvesting and sensing.[281, 282] Organic thermoelectric materials have also attracted research attention since they inherently have low thermal conductivity to start with.[283] With novel applications, we hope to see thermal conductivity to be at the driving seat for polymer innovations.

## 7. CONCLUDING REMARK

"*Any consideration of thermal conductivity, particularly in the context of amorphous polymers, is restricted by three principal factors. The first is the present lack of understanding of the nature of molecular organization in the amorphous, or so-called amorphous, state, whether a material is in the overt liquid or supercooled glassy phase. The second is the absence of any precise working theory for such a material, and the third is the paucity of data, coupled with some uncertainty in the reliability of even this. The latter represents the experimental difficulties associated with the measurement of thermal conductivity, although figures are usually available to two significant figures on most commercially available polymers.*" This is quoted from Hands' 1973 paper.[151] The first two restricting factors certainly remains true up to date, although accurately measuring polymer thermal conductivity can now be achieved using various techniques. The first factor is related to the chemistry-property relationships, while the second one calls for better models to describe the heat carrier transport physics. With this review, we hope to summarize the advancements in thermal transport physics in polymers and point out places where more research is needed. The underlying physics of thermal transport in crystalline fibers resemble a lot to that in inorganic lattices, but thermal transport in amorphous polymers is certainly a different scenario despite its similarity to amorphous inorganics. It is our belief that we are still far from completely understanding the complicated chemistry-property relationship for amorphous polymers. We hope there will be a collective effort from different communities to leverage different tools, including simulation, experiments and machine learning, to work towards better understanding polymer thermal transport.

**Acknowledgements:**

T.L. would like to acknowledge the support from the DuPont Young Professor Award, the Army Research Office (W911NF-16-1-0267) and the American Chemistry Society Petroleum Research Fund (54129-DNI10) on relevant research topics reviewed in this article. Z.T. acknowledge the support from 3M Non-Tenured Faculty Award and the American Chemistry Society Petroleum Research Fund (58688-DNI7). T.L.'s computation was supported in part by the University of Notre Dame, Center for Research Computing, and NSF through XSEDE (Extreme Science and Engineering Discovery Environment) resources provided by TACC Stampede-II under grant number TG-CTS100078. Z.T.'s work used the XSEDE with a grant number ACI-1053575.**References:**

1. Wong, C. in *Polymers for electronic & photonic application* (Elsevier, 2013).

2. Prasher, R. Thermal Interface Materials: Historical Perspective, Status, and Future Directions. *Proceedings of the IEEE* **94**, 1571-1586 (2006).

3. Chen, X., Su, Y., Reay, D. & Riffat, S. Recent research developments in polymer heat exchangers–A review. *Renewable and Sustainable Energy Reviews* **60**, 1367-1386 (2016).

4. Shen, S., Henry, A., Tong, J., Zheng, R. T. & Chen, G. Polyethylene nanofibres with very high thermal conductivities. *Nat Nanotechnol* **5**, 251-255 (2010).

5. Canetta, C., Guo, S. & Narayanaswamy, A. Measuring thermal conductivity of polystyrene nanowires using the dual-cantilever technique. *Rev. Sci. Instrum.* **85**, 104901 (2014).

6. Zhong, Z., Wingert, M. C., Strzalka, J., Wang, H., Sun, T., Wang, J., Chen, R. & Jiang, Z. Structure-induced enhancement of thermal conductivities in electrospun polymer nanofibers. *Nanoscale* **6**, 8283-8291 (2014).

7. Singh, V., Bougher, T. L., Weathers, A., Cai, Y., Bi, K., Pettes, M. T., McMenamin, S. A., Lv, W., Resler, D. P., Gattuso, T. R., Altman, D. H., Sandhage, K. H., Shi, L., Henry, A. & Cola, B. A. High thermal conductivity of chain-oriented amorphous polythiophene. *Nat Nano* **9**, 384-390 (2014).

8. Huang, X., Zhi, C., Jiang, P., Golberg, D., Bando, Y. & Tanaka, T. Polyhedral Oligosilsesquioxane-Modified Boron Nitride Nanotube Based Epoxy Nanocomposites: An Ideal Dielectric Material with HighThermal Transport in Polymers: A Review.                                                                74

115. Yao, Y., Zeng, X., Pan, G., Sun, J., Hu, J., Huang, Y., Sun, R., Xu, J. & Wong, C. Interfacial engineering of silicon carbide nanowire/cellulose microcrystal paper toward high thermal conductivity. *ACS Applied Materials & Interfaces* **8**, 31248-31255 (2016).

116. Yan, H., Mahanta, N. K., Majerus, L. J., Abramson, A. R. & Cakmak, M. Thermal conductivities of electrospun polyimide-mesophase pitch nanofibers and mats. *Polym. Eng. Sci.* **54**, 977-983 (2014).

117. Agrawal, A. & Satapathy, A. Development of a heat conduction model and investigation on thermal conductivity enhancement of AlN/epoxy composites. *Procedia Engineering* **51**, 573-578 (2013).

118. Guo, Y., Yang, X., Ruan, K., Kong, J., Dong, M., Zhang, J., Gu, J. & Guo, Z. Reduced Graphene Oxide Heterostructured Silver Nanoparticles Significantly Enhanced Thermal Conductivities in Hot-Pressed Electrospun Polyimide Nanocomposites. *ACS Appl. Mater. Interfaces* **11**, 25465-25473 (2019).

119. Wu, Y., Zhang, X., Negi, A., He, J., Hu, G., Tian, S. & Liu, J. Synergistic Effects of Boron Nitride (BN) Nanosheets and Silver (Ag) Nanoparticles on Thermal Conductivity and Electrical Properties of Epoxy Nanocomposites. *Polymers* **12**, 426 (2020).

120. Pang, Y., Yang, J., Curtis, T. E., Luo, S., Huang, D., Feng, Z., Morales-Ferreiro, J., Sapkota, P., Lei, F., Zhang, J., Zhang, Q., Lee, E., Huang, Y., Guo, R., Ptasinska, S., Roeder, R. K. & Luo, T. Exfoliated Graphene Leads to Exceptional Mechanical Properties of Polymer Composite Films. *ACS Nano* **13**, 1097-1106 (2019).

121. Ma, H. & Tian, Z. Chain rotation significantly reduces thermal conductivity of single-chain polymers. *J. Mater. Res.* **34**, 126-133 (2019).

122. Ma, H., Ma, Y. & Tian, Z. Simple Theoretical Model for Thermal Conductivity of Crystalline Polymers. *ACS Appl. Polym. Mater.* **1**, 2566-2570 (2019).

123. Bai, L., Zhao, X., Bao, R., Liu, Z., Yang, M. & Yang, W. Effect of temperature, crystallinity and molecular chain orientation on the thermal conductivity of polymers: a case study of PLLA. *J. Mater. Sci.* **53**, 10543-10553 (2018).

124. Muthaiah, R. & Garg, J. Temperature effects in the thermal conductivity of aligned amorphous polyethylene—A molecular dynamics study. *J. Appl. Phys.* **124**, 105102 (2018).

125. Zhang, Y., Zhang, X., Yang, L., Zhang, Q., Fitzgerald, M. L., Ueda, A., Chen, Y., Mu, R., Li, D. &